\documentclass[%
reprint,
 amsmath,amssymb,
 aps,
pre,
]{revtex4-2}

\usepackage{graphicx}% Include figure files
\usepackage{dcolumn}% Align table columns on decimal point
\usepackage{bm}% bold math
\usepackage{bigints}
\begin{document}

%\preprint{APS/123-QED}

\title{Partisan Voter Model: Stochastic description and noise-induced transitions}

\author{Jaume Llabr\'es}
 \affiliation{Institute for Cross-Disciplinary Physics and Complex Systems IFISC (CSIC-UIB), Campus UIB, 07122 Palma de Mallorca, Spain.}
\author{Maxi San Miguel}
 \affiliation{Institute for Cross-Disciplinary Physics and Complex Systems IFISC (CSIC-UIB), Campus UIB, 07122 Palma de Mallorca, Spain.}
\author{Ra\'ul Toral}
 \affiliation{Institute for Cross-Disciplinary Physics and Complex Systems IFISC (CSIC-UIB), Campus UIB, 07122 Palma de Mallorca, Spain.}

\date{\today}

\begin{abstract}
We give a comprehensive mean-field analysis of the Partisan Voter Model (PVM) and report analytical results for exit probabilities, fixation times, and the quasi-stationary distribution. In addition, and similarly to the noisy voter model, we introduce a noisy version of the PVM, named as the Noisy Partisan Voter Model (NPVM) which accounts for the preferences of each agent for the two possible states, as well as for idiosyncratic spontaneous changes of state. We find that the finite-size noise-induced transition of the noisy voter model is modified in the NPVM leading to the emergence of new intermediate phases and both continuous and discontinuous transitions. 
\end{abstract}

\maketitle

\section{\label{sec:introduction}Introduction}
The paradigmatic \textit{voter model}~\cite{liggett,Rednerreview} is a stochastic binary state model of opinion formation in a population of interacting agents that imitate each other at random. The imitation mechanism accounts for the herding phenomena observed in many social systems, and indeed this model has found wide applicability in the analysis of electoral processes~\cite{Fern_ndez_Gracia_2014}, language competition~\cite{Castell__2010}, amongst other, but also similar herding mechanisms have been identified in other distant fields such as biology, ecology, etc.~\cite{Clifford}. The model exhibits two absorbing states, known as consensus states, which represent situations in which all agents have adopted the same opinion. The standard voter model is characterized by the absence of free parameters, and therefore, lacks the capacity to describe the macroscopic transitions that occur in other models of collective social behavior for critical values of the parameters. 

A variation of the model, known as the \textit{noisy voter model}~\cite{kirman,CarroSciRep,peraltaNJP}, accounts for imperfect imitation in which agents can change their state spontaneously, independent of the state of the other agents. This idiosyncratic behavior prevents the existence of absorbing states and competes with the herding behavior. For a critical value of the parameter that measures the relative strengths of herding and idiosyncratic behaviors, there is a finite-size noise-induced transition from a state dominated by herding to a state dominated by idiosyncratic behavior. This transition shows up as a change in the relative maxima of the stationary probability distribution ~\cite{horsthemke1984noise,Toral:2011}. Similar transitions also appear in related models ~\cite{Fichthorn,Redner1989,Sole2022}. 

There are numerous other variants of the voter model, including those with nonlinear interactions \cite{qvoter,Schweitzer,nonlinearVM,PeraltaCHAOS}, effects of aging \cite{Artime_2018,Aging}, the presence of zealots \cite{Mobilia,zealotsVM}, multiple states \cite{lucia}, different preferences for the two possible states of agents~\cite{Czaplicka_2022,Masuda_2010,Masuda_2011}, and other modifications~\cite{Rednerreview}. In this paper we focus on the \textit{partisan voter model} (PVM)~\cite{Masuda_2010,Masuda_2011} in which every agent has a fixed preference for one of the two states. We also introduce a \textit{noisy partisan voter model} (NPVM), in which agents, in addition of a preferred state, exhibit idiosyncratic behavior with spontaneous change of state. 

While in the voter model the ensemble average of the proportion of agents in each state is a dynamically conserved quantity, in the partisan voter model the dynamics selects a state of the system which is not not determined by the initial proportion of agents in each state. The main question that we address is how the noise-induced transition of the noisy voter model is affected by the PVM's dynamically selected solution. We find that, in the symmetric case, where half of the agents prefer each state, the system undergoes a discontinuous transition where the probability distribution of the number of agents in one state changes from unimodal to trimodal. For a general proportion of agents preferring each state, we observe a rich phase diagram with continuous and discontinuous finite-size noise-induced transitions.

The paper is structured as follows: In Section~\ref{sec:rateeq} we review previous work of the partisan voter model~\cite{Masuda_2010,Masuda_2011} providing a comprehensive mean field theory. Additionally, in Section~\ref{sec:PVM_sto} we introduce a stochastic analysis of the model presenting new analytical results for the stationary probability distribution, exit probabilities, fixation times and quasi-stationary probability distribution. In Section~\ref{sec:NPVM}, we introduce the noisy partisan voter model and study it in the mean-field case, discussing the different noise-induced transitions. Finally, we conclude with some general remarks in Section~\ref{summary}.

\section{\label{sec:PVM} Partisan Voter Model}
We present the partisan voter model introduced in
~\cite{Masuda_2010,Masuda_2011}. The system consists of $N$ agents, ``voters'', connected by links. Throughout this work we limit ourselves to the all-to-all connected topology, or complete graph where each agent is connected to every other one. Voter $i\in[1,N]$ holds a binary state variable $s_i \in \{ -1, +1 \}$. This variable might have different meanings depending on the context, such as the language used by the speakers of a bilingual society or the voter's left or right political option, but its precise interpretation does not concern us in this paper. Agents can change their state by adopting the state of a randomly chosen neighbor but, at variance with the standard voter model, independently of its current state, every agent has an innate preference for one of the two states. Therefore, we can have four different types of agents that we label: $i^+_+$, $i^+_-$, $i^-_+$ and $i^-_-$, where the superscript indicates the preference and the subscript the state. The strength of the preference is quantified with the parameter $\varepsilon \in [0,1]$ and, although one can be more general~\cite{Masuda_2011}, we limit ourselves to the case in which the strength $\varepsilon$ is the same for all the voters. We denote by $q$ the fraction of agents that prefer to be in state $+1$.

The dynamics of the model is governed by the following rules: with a constant rate $h$ an agent, say $i$, randomly selects another agent, say $j$, from the set of all its neighbors. Once selected, there are two possible situations. If both agents are in the same state ($s_i=s_j$), nothing happens. If they are in different states, the agent $i$ changes state with a probability that depends on its preference, according to the following scenario:
\begin{itemize}
\item $i^+_+ \Rightarrow i^+_- $ with probability $\frac{1-\varepsilon}{2}$,
\item $i^+_- \Rightarrow i^+_+ $ with probability $\frac{1+\varepsilon}{2}$,
\item $i^-_- \Rightarrow i^-_+ $ with probability $\frac{1-\varepsilon}{2}$,
\item $i^-_+ \Rightarrow i^-_- $ with probability $\frac{1+\varepsilon}{2}$.
\end{itemize}

If $\varepsilon=0$, we recover the updating rules of the voter model, in which the agent copies a neighbor with probability $1/2$. On the other hand, for $\varepsilon=1$, the agent is a zealot that does not change its state when it is aligned with its preference. This latter extreme case will not be covered in this work and, instead, we have focused on small values of $\varepsilon>0$. The system also bears some similarities with the {\slshape biased voter model} where only a fraction of the population is biased towards one of the two options, while the rest of the population is neutral~\cite{Czaplicka_2022}. As in the standard voter model, the system may enter into an absorbing configuration from which no further evolution is possible. The PVM presents two absorbing states corresponding to the two configurations in which all agents are in the same state, either $+1$ or $-1$. These are also known as consensus states.
\subsection{Rate equations and dynamical system}\label{sec:rateeq}
\begin{figure*}
\centering
\includegraphics[width=\textwidth]{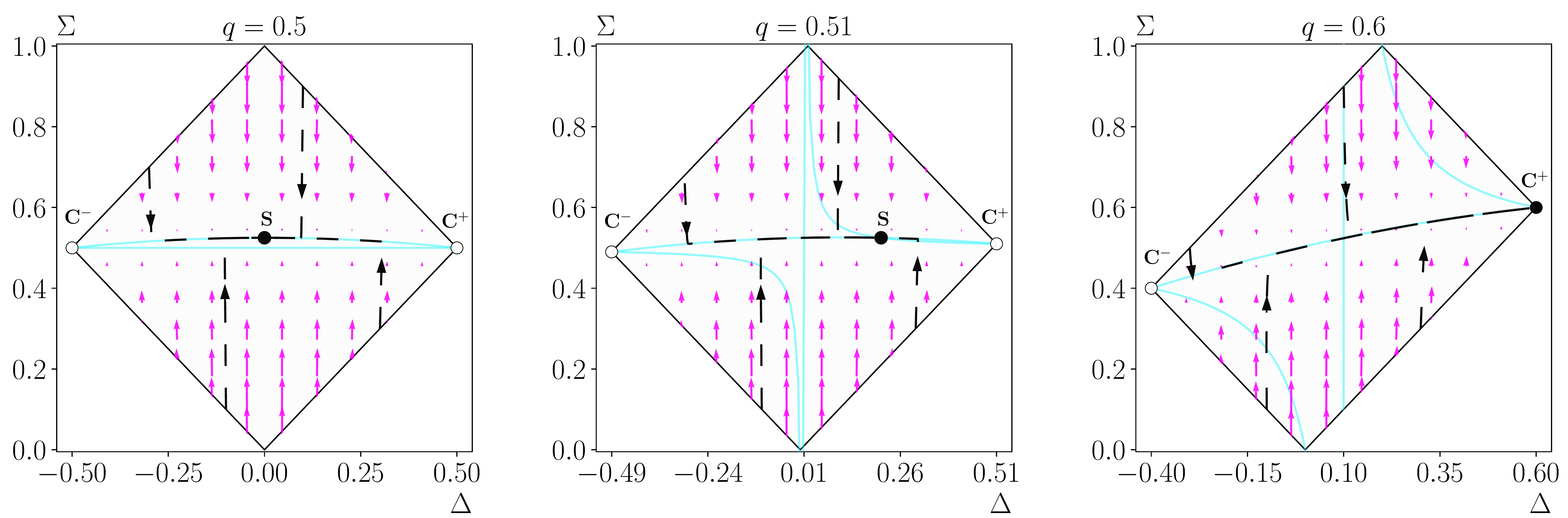}
\caption{Flow diagram in the phase space $(\Delta,\Sigma)$ for $\varepsilon=0.05$ and $q=0.5, 0.51, 0.6$ as indicated. The critical density of agents is $q_c^+=0.525$. The rectangle shows the allowed locations in the plane defined by Eqs.~\eqref{eq:squareregion}. Arrows indicate the direction of the velocity field. The larger the arrow, the greater the velocity. The filled (resp. open) symbol corresponds to a stable (resp. saddle) fixed point. The blue lines indicate the nullclines given by Eqs.~(\ref{eq:PVM_nullcline_delta}, \ref{eq:PVM_nullcline_sigma}). The black dashed lines display deterministic trajectories, starting from different locations at the boundaries of the allowed region, obtained by a numerical integration of Eqs.~(\ref{eq:dots1},\ref{eq:dots2}) using the forward Euler method with an appropriate time step.}
\label{fig:PVM_DS}
\end{figure*}

We first perform a deterministic analysis of the rate equations and the dynamical system in the mean-field limit, suitable to the all-to-all connected topology in the thermodynamic limit $N\to\infty$. Let $x^a_b=n^a_b/N$ be the density of $i^a_b$ type agents. Considering the different possible interactions between agents in opposite states, the rate equation for the density of agents $x_+^+$ reads \cite{Masuda_2010}
\begin{equation}\begin{split}\label{x++}
\frac{d x^+_+}{d t}=&x^+_- x^+_+(1+\varepsilon) + x^+_- x^-_+ \left(1+\varepsilon\right) \\ &- x^+_+ x^-_- \left(1-\varepsilon\right)-x^+_+ x^+_-(1-\varepsilon),
\end{split} 
\end{equation}
where, without loss of generality, we have rescaled time to dimensionless units as $t\to\frac{h}{2}t$. In the same way, one can write down the rate equations for the densities $x_+^-$, $x_-^+$ and $x_-^-$. The conservation of the number of agents: $x_+^++x_+^-+x_-^++x_-^-=1$ and the fact that the preference is fixed, i.e. $x^+_++x^+_-=q$ and $x^-_-+x^-_+=1-q$, allow us to describe the system with just two independent variables. For convenience, they have chosen to be the difference $\Delta\equiv x^+_+ - x^-_- $ and the sum $\Sigma\equiv x^+_+ + x^-_- $ of the densities of the voters that are in their preferred state. $\Sigma$ can then be interpreted as the density of ``satisfied'' agents, those whose state coincides with their preference, while $\Delta$ is directly related with the {\slshape magnetization} $m=x^+_++x^-_+-x^+_--x^-_-$ as $m=1-2q+2\Delta$. Due to their definition, it is easy to see that not all the locations in the $(\Delta,\Sigma)$ plane are allowed and that the values of these variables are bounded to the rectangular area defined by
\begin{equation}\label{eq:squareregion}
\begin{split}
0\le \Sigma&\le 1,\\
\Sigma-2(1-q)\le \Delta& \leq 2q-\Sigma, \\
-\Sigma\le \Delta & \le\Sigma,\\
q-1\le \Delta&\le q,
\end{split}
\end{equation}
as shown in Fig.~\ref{fig:PVM_DS}.

Using the evolution equation Eq.~\eqref{x++} and similar ones for $x_+^-$, $x_-^+$ and $x_-^-$, one can derive the rate equations of $\Delta$ and $\Sigma$~\cite{Masuda_2010,Masuda_2011}
\begin{eqnarray} \label{eq:dots1}
\dot{\Delta}&=&\varepsilon \Delta - 2\varepsilon \Delta\Sigma+\varepsilon\Sigma(2q-1),\\ 
\dot{\Sigma}&=&2q(1 + \varepsilon)(1-q) \nonumber \\ &&+\Delta(1+2\varepsilon)(2q-1)- \Sigma - 2\varepsilon \Delta^2.\label{eq:dots2}
\end{eqnarray}
Note that for $\varepsilon=0$, only one variable is needed to describe the system as $\Delta$ is a constant given by the initial conditions. 

For general $0<\varepsilon<1$, the two consensus absorbing states are fixed points of the dynamical equations: 
\begin{itemize}
\item \textit{Consensus}$^-$($\mathrm{C}^-$): $\left(\Delta,\Sigma\right)=\left(q-1,1-q\right)$. All the voters are in the state $-1$. 
\item \textit{Consensus}$^+$($\mathrm{C}^+$): $\left(\Delta,\Sigma\right)=\left(q,q\right)$. All the voters are in the state $+1$. 
\end{itemize}
Note that once one of these two absorbing fixed points has been reached, the stochastic dynamics is not able to exit from it.

A third fixed-point solution is possible,
\begin{itemize}
\item \textit{Self-Centered}(S):\\ $\left(\Delta,\Sigma\right)=\left(\Delta^*,\Sigma^*\right)\equiv\left(\frac{\left(1+\varepsilon\right)\left(2q-1\right)}{2\varepsilon},\frac{1+\varepsilon}{2}\right)$. 
\end{itemize}
However, this solution only fulfills the boundary conditions given by Eqs.~\eqref{eq:squareregion} when $q$ satisfies 
\begin{equation} \label{eq:PVM_critical_qs}
\frac{1-\varepsilon}{2}\equiv q_c^-\le q\le q_c^+\equiv \frac{1+\varepsilon}{2}.
\end{equation}
 The self-centered solution represents a coexistence situation in which agents in both states are present and a majority of voters hold a state that agrees with their preference. At variance with the consensus solutions $\mathrm{C}^\pm$, the stochastic rules allow the system to exit the self-centered solution, as it is non-absorbing. For $q=q_c^-$ (resp. $q=q_c^+$), the self-centered coincides with the C$^-$ (resp. C$^+$) consensus state.

The stability of the fixed points can be found by means of a linear stability analysis. If $q_c^-\le q\le q_c^+$ the self-centered solution is stable, while the consensus solutions are unstable saddle points. If $q\le q_c^-$ (resp. $q\ge q_c^+$) the fixed point $\mathrm{C}^-$ (resp. $\mathrm{C}^+$) becomes stable. 

The nullclines of Eqs.~(\ref{eq:dots1}, \ref{eq:dots2}), $\dot{\Delta}=~0$ and $\dot{\Sigma}=0$ are, respectively \cite{Masuda_2011}: 
\begin{eqnarray} \label{eq:PVM_nullcline_delta}
\Sigma &=& \frac{\Delta}{1+2\Delta-2q},\\
\Sigma&=&- 2\varepsilon \Delta^2+2q\left(1 + \varepsilon\right)\left(1-q\right) 
\label{eq:PVM_nullcline_sigma}\\
\nonumber
&&+\Delta\left(1+2\varepsilon\right)\left(2q-1\right)
\end{eqnarray}
In Fig.~\ref{fig:PVM_DS}, we display the phase space $(\Delta,\Sigma)$ for three different values of $q$ together with some deterministic trajectories. The fact that the time derivative $\dot \Delta$ is proportional to the preference $\varepsilon$, making it a slow variable in the small $\varepsilon$ limit, has interesting consequences. As can be observed from the numerical integration, trajectories tend toward the only stable solution (either the self-centered point for $q_c^-\le q\le q_c^+$ or one of the consensus points otherwise), in two steps: First, the variable $\Sigma(t)$ quickly evolves while $\Delta(t)$ remains practically constant, until the nullcline $\dot{\Sigma}=0$, Eq.~\eqref{eq:PVM_nullcline_sigma}, is reached. Afterwards the trajectory follows closely that nullcline until the fixed point is reached asymptotically. In this second stage, one can obtain an accurate description by slaving the evolution of $\Sigma(t)$ to that of $\Delta(t)$ and write down a rate equation for the slow variable $\Delta(t)$ after substituting Eq.~\eqref{eq:PVM_nullcline_sigma} into Eq.~\eqref{eq:dots1} as
\begin{equation}\label{eq:PVM_dots_reduced}
\dot{\Delta}=2 \varepsilon (\Delta - q) (1 + \Delta - q) \left((1 + \varepsilon)(1 - 2 q)+ 2 \varepsilon \Delta\right).
\end{equation}
The accuracy of this adiabatic elimination increases as the strength of the preference $\varepsilon$ decreases. 

Although we can reach an understanding of the dynamics or the model from the deterministic analysis, we must emphasize that from the stochastic point of view the behavior will be rather different. Despite the existence of the stable self-centered fixed point, the stochastic dynamics of the system will inevitably reach one of the absorbing states. In the next section, we will describe several relevant features of the stochastic dynamics.

\subsection{Stochastic analysis}\label{sec:PVM_sto}

\begin{figure*}
\centering
\includegraphics[width=\textwidth]{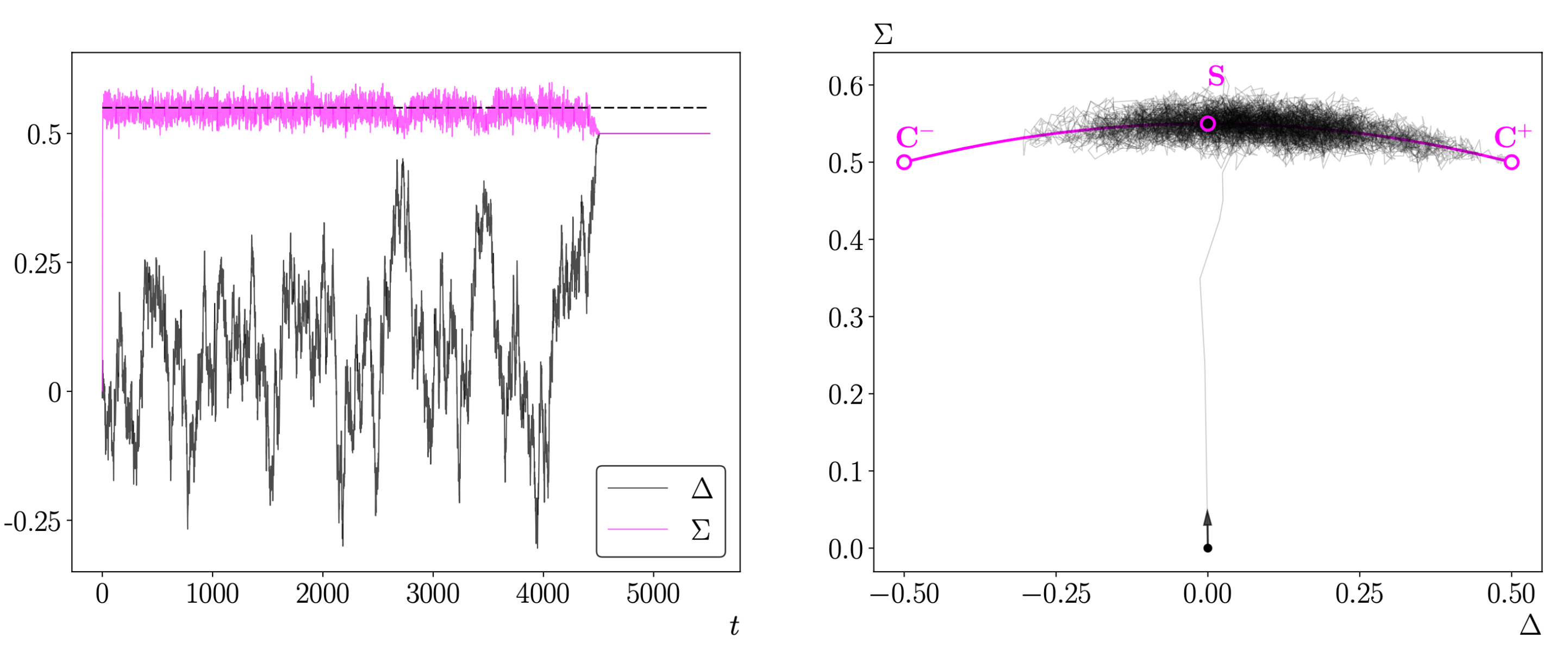}
\caption{Plot of a typical trajectory of the stochastic partisan voter model defined by the rates Eq.~\eqref{eq:PVM_globalrates}. Starting point is $\left(\Delta,\Sigma\right)=\left(0,0\right)$. Left panel: time evolution of both $\Delta$ and $\Sigma$. Dotted black line indicates the value of the self-centered solution, $\left(\Delta^*,\Sigma^*\right)=\left(0,0.55\right)$. Right panel: stochastic trajectory in the phase space $(\Delta,\Sigma)$. Darker areas indicate where the system stays for longer time. Magenta line represents the nullcline $\dot{\Sigma}=0$ of Eq.~\eqref{eq:PVM_nullcline_sigma}. Parameter values: $N=1000$, $\varepsilon=0.10$, $q=0.50$. All the results of computer simulations shown in this paper have been performed using the Gillespie algorithm~\cite{Gillespie:1977db, Toral2014StochasticNM}. }
\label{fig:PVM_stochastic_trajectory}
\end{figure*}

Let $W^{\pm\pm}\equiv W\left(\Delta \to \Delta \pm 1/N;\Sigma \to \Sigma \pm 1/N \right)$ be the rates at which the global variables $\Delta$ and $\Sigma$ evolve. Taking into account the possible interactions between agents in the all-to-all connected topology, those global rates are given by
\begin{equation} \label{eq:PVM_globalrates}
\begin{split}
W^{++}&=\frac{Nh}{2}\left(1-q+\Delta\right)(2q-\Delta-\Sigma)\left(\frac{1+\varepsilon}{2}\right), \\
W^{+-}&=\frac{Nh}{2}\left(1-q+\Delta\right)(\Sigma-\Delta)\left(\frac{1-\varepsilon}{2}\right), \\
W^{-+}&=\frac{Nh}{2}\left(q-\Delta\right)(2-2q+\Delta-\Sigma)\left(\frac{1+\varepsilon}{2}\right), \\
W^{--}&=\frac{Nh}{2}\left(q-\Delta\right)(\Sigma+\Delta)\left(\frac{1-\varepsilon}{2}\right).
\end{split}
\end{equation}
In Fig.~\ref{fig:PVM_stochastic_trajectory} we plot one stochastic trajectory in the symmetric case $q=\frac{1}{2}$ and $\varepsilon=0.1$. Despite the stochastic behavior, we can identify similarities with the deterministic trajectories of Fig.~\ref{fig:PVM_DS}. Although the trajectory is attracted toward the self-centered fixed point $\left(\Delta^*,\Sigma^*\right)$, it fluctuates along the nullcline $\dot{\Sigma}=0$ until eventually falls into the absorbing state $\mathrm{C}^{+}$. Note that the fast variable $\Sigma$ fluctuates slightly around $\Sigma^*$, while the slow variable $\Delta$ performs much larger fluctuations before both of them reach the absorbing state $\text{C}^+$.

Using standard techniques~\cite{vanKampen:2007,Toral2014StochasticNM}, one can write down the corresponding master equation of the process for the probability $ P(\Delta, \Sigma;t)$ of the values $\Delta$ and $\Sigma$ at a time $t$ using the rates of Eq.~\eqref{eq:PVM_globalrates}. However, the resulting equation is too complicated and we have not been able to find its solution even in the stationary situation. In order to make progress, we will use the adiabatic elimination technique~\cite{haken1983synergetics}, which applies when the variables of a dynamical system present different time scales as it is the case here for sufficiently small preference $\varepsilon$. The aim is to describe approximately the behavior of the system at long time scales by eliminating the fast variable that quickly reaches a stationary value associated to a given value of the slow variable. 

Here, the adiabatic elimination is implemented using a reduction method proposed in~\cite{Pineda2009}. Let us write the joint probability density function of the system taking values $\Delta$ and $\Sigma$ at time $t$ as
\begin{equation}
P\left(\Delta, \Sigma ;t\right)=P\left(\Delta;t\right)\hspace{0.03cm} P\left(\Sigma \hspace{0.02cm}|\hspace{0.02cm}\Delta ;t\right),
\end{equation}
where $P\left(\Sigma \hspace{0.02cm}|\hspace{0.02cm}\Delta ;t\right)$ is the conditional probability density to obtain $\Sigma$ given $\Delta$ at a time $t$ and $P\left(\Delta;t\right)$ is the probability density to obtain $\Delta$ at the same time. Due to the dynamics of the fast variable, the conditional probability distribution quickly tends to a narrow and sharp curve centered on the value given by the nullcline $\dot{\Sigma}=0$. Therefore, $\Sigma$ is directly determined from $\Delta$ and one can write down an effective stochastic dynamics only taking into consideration the variable $\Delta$. Substituting Eq.~\eqref{eq:PVM_nullcline_sigma} into Eqs.~\eqref{eq:PVM_globalrates} one can write down the effective global rates of this stochastic dynamics as
\begin{equation} \label{eq:reducedglobalrates}
\begin{split}
W^+(\Delta)\equiv& \,W^{++}+ W^{+-}\\ 
=&\,\frac{N}{2}\left(q-\Delta\right)\left(1-q+\Delta\right)\\
&\,\times \left(1-\varepsilon-2\varepsilon^2(1+\Delta)+2\varepsilon q(1+\varepsilon)\right), \\
W^-(\Delta) \equiv&\, W^{-+}+ W^{--}\\ 
=&\,\frac{N}{2}\left(q-\Delta\right)\left(1-q+\Delta\right)\\
&\,\times\left(1+\varepsilon+2\Delta\varepsilon^2-2\varepsilon q(1+\varepsilon)\right).
\end{split}
\end{equation}
Starting from those rates one can write down the Fokker-Planck equation of the reduced model using a standard approach~\cite{vanKampen:2007,Toral2014StochasticNM}, 
\begin{equation} \label{eq:FP_1D}
\frac{\partial P(\Delta;t)}{\partial t}=\mathbf{H} P(\Delta;t),
\end{equation}
where the operator $\mathbf{H}$ is defined as
\begin{equation} \label{eq:adjointH}
\mathbf{H}=-\frac{\partial}{\partial \Delta}F(\Delta)+\frac{\partial^2}{\partial \Delta^2}D(\Delta).
\end{equation}
The functions $F(\Delta)=\left(W^+(\Delta)-W^-(\Delta)\right)/N$ and $D(\Delta)=~\left(W^+(\Delta)+W^-(\Delta)\right)/N^2$ are commonly called drift and diffusion, respectively, and, for this model, are given by
\begin{eqnarray} \label{eq:PVM_FPterms}
F(\Delta)&=&-\varepsilon\left(q-\Delta\right)\left(1-q+\Delta\right)\left(2\Delta\varepsilon-\left(1-2q\right)(1+\varepsilon)\right), \nonumber \\
D(\Delta)&=&\frac{1}{2N}\left(1-\varepsilon^2\right)\left(q-\Delta\right)\left(1-q+\Delta\right).
\end{eqnarray}

\begin{figure*}[t]
\centering
\includegraphics[width=\textwidth]{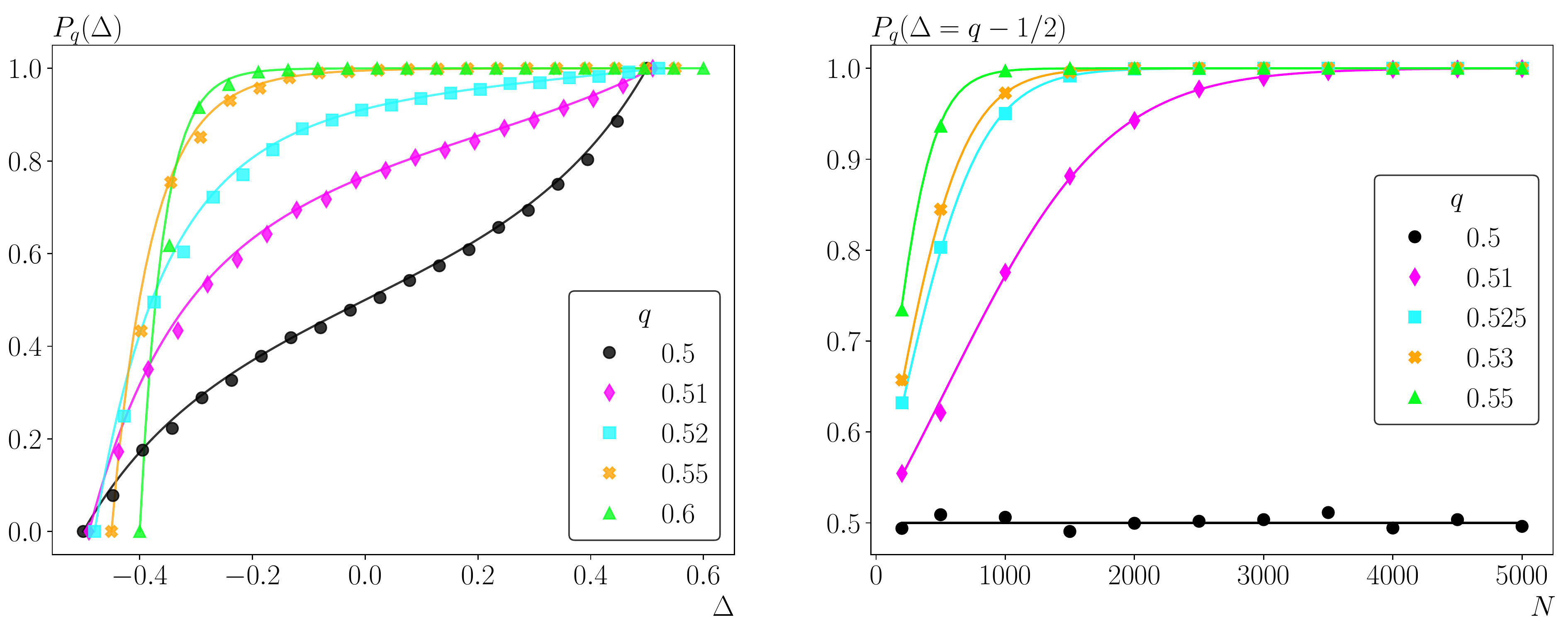}
\caption{Left panel: Probability $P_{q}(\Delta)$ to reach the consensus state $+1$, as a function of the initial condition $\Delta$ for different fraction of agents $q=0.5, 0.51, 0.52, 0.55, 0.6$ as indicated for a system size $N=1000$. Right panel: $P_{q}(\Delta=q-\frac{1}{2})$ versus the system size for different fraction of agents $q=0.5, 0.51, 0.525, 0.53, 0.55$ as indicated. In both panels the preference is $\varepsilon=0.05$. Solid lines correspond to the analytical result given by Eq.~\eqref{eq:PVM_PexitQ} while symbols indicate the results of computer simulations of the complete model.}
\label{fig:PVM_Pexit}
\end{figure*}

We limit ourselves to analyze the stationary distribution $P_{\mathrm{st}}(\Delta)$, which can be expressed, after setting the left hand side to 0 in Eq.~\eqref{eq:FP_1D}, as an exponential function,
\begin{equation} \label{eq:Pstform}
 P_{\mathrm{st}}(\Delta)=\mathcal{Z}^{-1}\cdot\exp{\left[-V(\Delta)\right]},
\end{equation}
where $\mathcal{Z}$ is the normalization constant and $V(\Delta)$ is a ``potential function" given by
\begin{equation} \label{eq:potential}
 V(\Delta)=\int^\Delta\frac{-F(z)+D'(z)}{D(z)}dz.
\end{equation}
In our particular case, the stationary distribution reads
\begin{equation}\label{eq:ABD_Pst}
P_{\mathrm{st}}(\Delta)=\mathcal{Z}^{-1} \frac{\mathrm{exp}\left[-\frac{\beta}{2}\Delta^2-\gamma\Delta\right]}{(q-\Delta)(1-q+\Delta)},
\end{equation}
 where the parameters $\beta$ and $\gamma$ are given by
\begin{equation}
\begin{split}
\beta&=\frac{4N\varepsilon^2}{1-\varepsilon^2}, \\ \gamma&=\frac{2N\varepsilon}{1-\varepsilon}\left(1-2q\right) ,
\end{split}
\end{equation}
and where the normalization constant $\mathcal{Z}$ has to be determined with the condition $\int_{q-1}^{q} P_{\mathrm{st}}(\Delta)d\Delta=1$. However, this integral diverges in both limits $\Delta=q-1$ and $\Delta=q$ and therefore the stationary distribution can not be normalized. We interpret this result as follows: consensus configurations are the only absorbing states and they are reached with probability one in a finite-size system. The probability of reaching one or the other consensus state depends on the initial condition $\Delta_0$. Hence, the stationary probability distribution must include a sum of Dirac-delta functions in the two absorbing states weighted with the probability to reach each one of them, 
\begin{equation} \label{eq:PVM_Pexit_delta}
\begin{split}
P_{st}\left(\Delta| \Delta_0\right)=&P_q(\Delta_0)\delta\left(\Delta-q\right)\\
&+(1-P_{q}(\Delta_0))\delta\left(\Delta-q+1\right),
\end{split}
\end{equation}
where $P_{q}(\Delta_0)$ is the exit (or fixation) probability, defined as the probability that a finite system with an initial configuration $\Delta_0$ reaches a consensus to the absorbing state $\Delta=q$. In the following, we study the exit probability in order to determine the weights of each Dirac-delta of Eq.~\eqref{eq:PVM_Pexit_delta}.

\subsubsection{Exit probability}
The probability $P_{q}(\Delta)$ of reaching a consensus on the absorbing state $+1$ starting from an initial condition $\Delta_0=\Delta$ can be shown to satisfy the recurrence relation~\cite{redner_2001}
\begin{equation} \label{eq:PVM_Pexit1}
\begin{split}
&\left(W^+(\Delta)+W^-(\Delta)\right)P_{q}(\Delta)\\
&=W^+ (\Delta)P_{q}\left(\Delta+1/N\right)+W^- (\Delta)P_{q}\left(\Delta-1/N\right)
\end{split}
\end{equation}
with boundary conditions $P_{q}\left(q-1\right)=0$ and $P_{q}\left(q\right)=1$. Although a rigorous treatment of this recursion relation is rather complicated, an approximate solution can be obtained by expanding Eq.~\eqref{eq:PVM_Pexit1} up to second order in $1/N$, leading to
\begin{equation}\label{eq:PVM_Pexit_2}
\frac{d^2P_{q}}{d\Delta^2}=\left(\beta\Delta+\gamma\right)\frac{dP_{q}}{d\Delta},
\end{equation}
with the same boundary conditions. Note that this equation can also be written as
\begin{equation}\label{eq:P_H}
\mathbf{H^+}P_q=0,
\end{equation}
where $\mathbf{H^+}$ is the adjoint operator of the Fokker-Planck equation, Eqs.~(\ref{eq:FP_1D}, \ref{eq:adjointH}),
\begin{equation}
\mathbf{H^+}=F(\Delta)\frac{\partial}{\partial \Delta}+D(\Delta)\frac{\partial^2}{\partial \Delta^2}.
\end{equation}

The solution of Eq.~\eqref{eq:PVM_Pexit_2} for the aforementioned boundary conditions is given by
\begin{equation} \label{eq:PVM_PexitQ}
P_{q}\left(\Delta\right)=\frac{\mathrm{erfi}\left(\sqrt{\frac{\beta}{2}}\left(\Delta+\frac{\gamma}{\beta}\right)\right) - \mathrm{erfi}\left(\sqrt{\frac{\beta}{2}}\left(q-1+\frac{\gamma}{\beta}\right)\right)}{\mathrm{erfi}\left(\sqrt{\frac{\beta}{2}}\left(q+\frac{\gamma}{\beta}\right)\right) - 
{\mathrm{erfi}\left(\sqrt{\frac{\beta}{2}}\left(q-1+\frac{\gamma}{\beta}\right)\right)}},
\end{equation}
where $\mathrm{erfi(\cdot)}$ is the imaginary error function defined as $\mathrm{erfi}(z)=-i\, \mathrm{erf}(iz)$ in terms of the standard error function $\mathrm{erf}(\cdot)$. Specifically, in the symmetric case $q=\frac{1}{2}$, Eq.~\eqref{eq:PVM_PexitQ} takes the following form: 
\begin{equation}\label{eq:PVM_Pexit_sol_symm}
P_{1/2}(\Delta)=\frac{1}{2}\left[1+\frac{\mathrm{erfi}\left(\sqrt{\frac{\beta}{2}}\Delta\right)}{\mathrm{erfi}\left(\frac{1}{2}\sqrt{\frac{\beta}{2}} \right)}\right].
\end{equation}
Note that for $\beta=0$ ($\varepsilon=0$) we recover the expression of the standard voter model, $P_{1/2}\left(\Delta\right)=\frac{1}{2}+\Delta=\sigma$, with $\sigma$ the density of voters in the state~$+1$.

In the left panel of Fig.~\ref{fig:PVM_Pexit} we display the exit probability $P_{q}\left(\Delta\right)$ as a function of the initial condition $\Delta$. The results of the numerical simulations of the model agree very well with the analytical solution given by Eq.~\eqref{eq:PVM_PexitQ}. This indicates that the adiabatic elimination technique is a reliable approximation for the small value $\varepsilon=0.05$ taken in the figure. Furthermore, in Appendix~\ref{app:PVM_exitp_complete} we show that the adiabatic approximation matches well the results coming from a numerical solution of the master equation of the complete model without the adiabatic elimination. Regarding the asymmetric case $q\neq\frac{1}{2}$, a slight increase in $q$ ($q=0.51$ or $q=0.52$) significantly increases the exit probability. When $q=0.60$, the system predominantly converges to the consensus state $+1$ for a vast majority of the initial conditions. In fact, in this case, the dependence of the fixation probability with the initial magnetization is very similar to the one found in the biased voter model~\cite{Czaplicka_2022}. Focusing on the symmetric case ($q=\frac{1}{2}$, or $\gamma=0$), we notice that the slope of $P_{q}\left(\Delta\right)$ is higher near the absorbing states than in the rest of the interval. At the center of the interval $\Delta=0$, the exit probability is $\frac{1}{2}$. In the right panel of Fig.~\ref{fig:PVM_Pexit}, we plot the exit probability versus the system size $N$ while the preference $\varepsilon$ is fixed for the initial condition $\Delta=q-\frac{1}{2}$. In the symmetric case, the fixation probability remains at $\frac{1}{2}$ while in the asymmetric case the exit probability tends to $1$. 

\begin{figure}[b]
\centering
\includegraphics[width = \columnwidth]{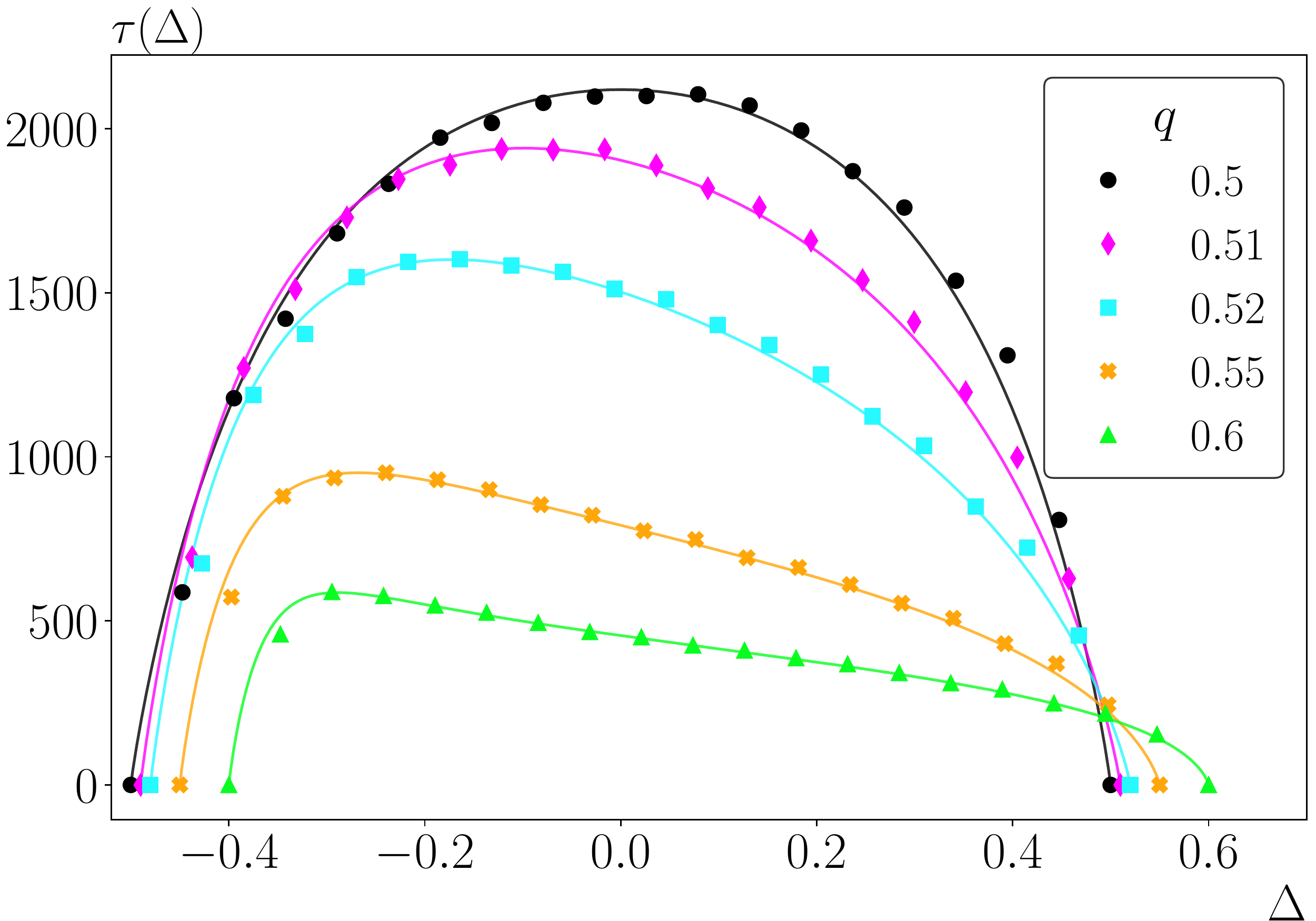}
\caption{Plot of the average time $\tau$ to reach an absorbing state, in units of MCS, as a function of the initial condition $\Delta$ for different fraction of agents $q=0.50,\, 0.51,\, 0.52,\, 0.55,\, 0.6$ as indicated. Solid lines correspond to the analytical result given by Eq.~\eqref{eq:PVM_TC_sol} while symbols indicate the results of computer simulations of the complete model. Parameter values: $N=~1000$, $\varepsilon=0.05$.}
\label{fig:PVM_tau_deltas}
\end{figure}

\begin{figure*} [t] 
\includegraphics[width = \textwidth]{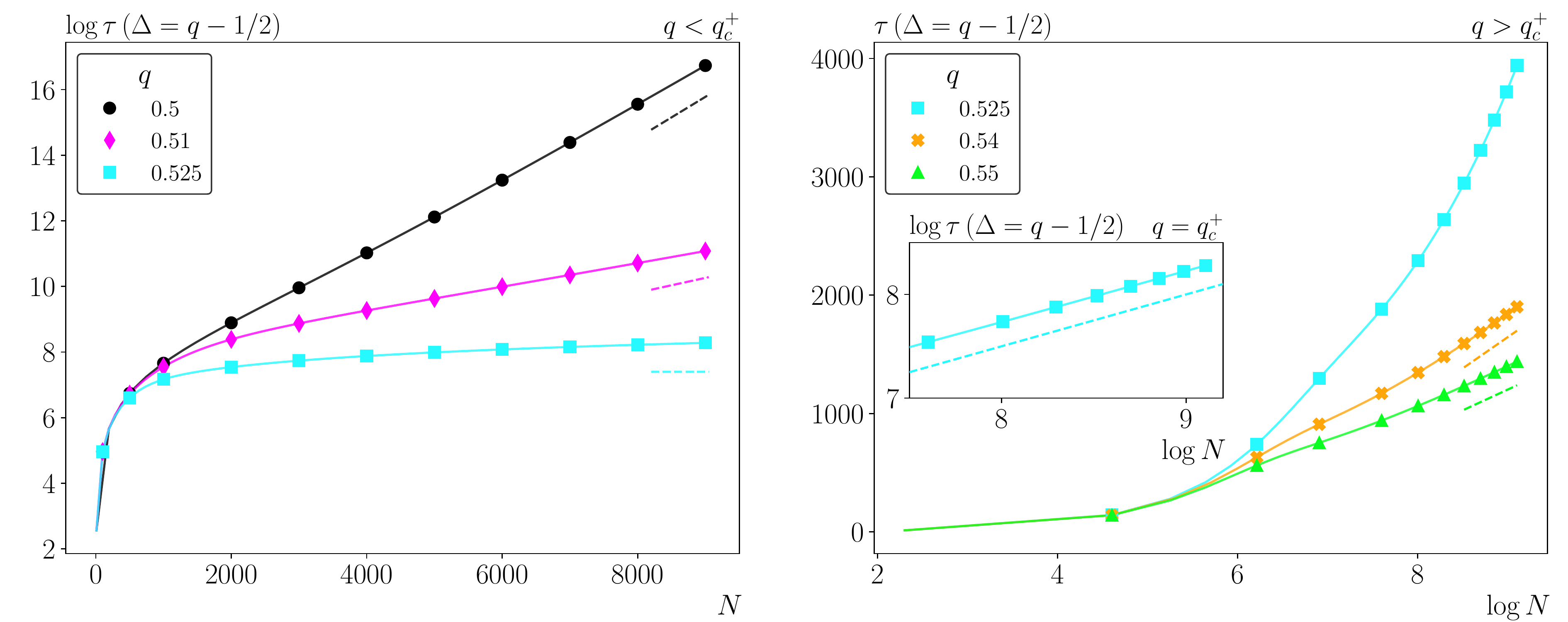}
\caption{\label{fig:tau_sizes} Left panel: Plot of the logarithm of the average time $\tau$ to reach an absorbing state, in units of MCS, versus the system size $N$ for $q=0.5,\,0.51,\,0.525 (=q_c^+)$ as indicated. Short dashed lines correspond to the exponential dependence of $\tau$ with $N$ given by Eq.~\eqref{eq:PVM_TClim}. Right panel: Plot of the the average time $\tau$ to reach an absorbing state, in units of MCS, versus the logarithm of the system size $N$ for $q=0.525 (=q_c^+),\,0.53,\,0.55$ as indicated. Short dashed lines indicates the best linear fit to the last points of the curve. Inset: Log-log plot of $\tau$ versus $N$ for $q=q_c^+=0.525$. Dashed line corresponds to a straight line with a slope $\frac{1}{2}$. In both panels, solid lines correspond to the solution of Eq.~\eqref{eq:PVM_TC_aprr} explicitly given by Eqs.~(\ref{eq:PVM_tau_sol}, \ref{eq:PVM_TC_sol}) while symbols indicate the results of computer simulations of the complete model. Parameter values: $\varepsilon=0.05$, $\Delta=q-\frac{1}{2}$.}
\end{figure*}

In principle, replacement of Eq.~\eqref{eq:PVM_PexitQ} in Eq.~\eqref{eq:PVM_Pexit_delta} leads to the determination of the stationary distribution. However, a word of caution is required concerning the situation in the thermodynamic limit $N\to\infty$. As, in this limit, Eq.~\eqref{eq:PVM_PexitQ} becomes independent of the initial condition,
\begin{equation}\label{PVM_limit_exitP}
\begin{split}
\lim_{N\to\infty}&P_q\left(\Delta\right)
= \left\{ \begin{array}{lcc} 0, &\mathrm{if} & q < \frac{1}{2}, \\ \\ \frac{1}{2}, & \mathrm{if} & q=\frac{1}{2}, \\\\ 1, & \mathrm{if} & q > \frac{1}{2}, \end{array}
\right.
\end{split}
\end{equation}
we obtain
\begin{equation}\label{PVM_limitstP}
\begin{split}
&\lim_{N\to\infty}\lim_{t\to\infty}P\left(\Delta;t\right) \\
&=\left\{ \begin{array}{lcc} \delta\left(\Delta+1-q\right), &\mathrm{if} & q < \frac{1}{2}, \\ \\ \frac{1}{2}\left(\delta\left(\Delta-\frac{1}{2}\right)+ \delta\left(\Delta+\frac{1}{2}\right)\right), & \mathrm{if} & q = \frac{1}{2}, \\\\ \delta\left(\Delta-q\right), & \mathrm{if} & q > \frac{1}{2}. \end{array}
\right.
\end{split}
\end{equation}

This result implies that the system will surely reach the most favored absorbing state, while in the symmetric case, it is equiprobable to end up in any of the absorbing states. This seems to be in contradiction with the mean-field analysis, that states that the final state of the system is the only stable fixed point, including the self-centered solution when available, and the fact that the mean-field description is valid in the thermodynamic limit. The answer to this contradiction lies on the fact that the stationary distribution in the mean-field description is obtained taking first the $N\to\infty$ limit and then the $t\to\infty$ limit, and that these two limits do not commute. Taking the infinite-time limit first, we obtain Eq.~\eqref{PVM_limitstP}, while taking the thermodynamic limit first, the stationary probability distribution reads

\begin{equation}\label{PVM_limitsNP}
\begin{split}
&\lim_{t\to\infty}\lim_{N\to\infty}P\left(\Delta;t\right) \\
&= \left\{ \begin{array}{lcc} \delta\left(\Delta+1-q\right), &\mathrm{if} & q \leq q_c^-, \\ \\ \delta\left(\Delta-\Delta^*\right), & \mathrm{if} & q_c^-\leq q \leq q_c^+, \\\\ \delta\left(\Delta-q\right), & \mathrm{if} & q \geq q_c^+. \end{array}
\right.
\end{split}
\end{equation}
We then recover the mean-field result: the system always ends up in the stable fixed point. The location of the fixed point depends on the value of $q$: self-centered if $q_c^-\leq q \leq~q_c^+$, or the corresponding consensus state otherwise.

In the following, we determine the average time a finite system needs to reach one of the absorbing states.

\subsubsection{Time to reach consensus}
Given an initial condition $\Delta_0=\Delta$, $\tau(\Delta)$ is defined as the average time to reach any of the two possible consensus states. This time can be obtained from the transition rates by means of the relation~\cite{redner_2001}
\begin{equation} \label{eq:PVM_TC}
\begin{split}
W^+(\Delta)&\tau\left(\Delta+1/N\right)
+W^-(\Delta)\tau\left(\Delta-1/N\right) \\ 
&-\left[W^+(\Delta)+W^-(\Delta)\right] 
\tau\left(\Delta\right)=-1,
\end{split}
\end{equation}
with boundary conditions $\tau\left(q-1\right)=\tau\left(q\right)=0$. 

As before, an approximate solution of this recurrence relation can be obtained by expanding it up to the second order in $1/N$, arriving at 
\begin{equation}\label{eq:PVM_TC_aprr}
\begin{split}
\frac{d^2\tau}{d\Delta^2}=&
\left(\beta \Delta+ \gamma \right)\frac{d\tau}{d\Delta}\\
&-\frac{2N}{1-\varepsilon^2}\frac{1}{\left(q-\Delta\right)\left(\Delta-(q-1)\right)},
\end{split}
\end{equation}
with the same boundary conditions. This equation can also be obtained using the adjoint operator of the Fokker-Planck equation $\mathbf{H^+}$ (Eq.~(\ref{eq:adjointH})) \cite{gardiner}
\begin{equation}\label{eq:TC_H}
\mathbf{H^+}\tau=-1.
\end{equation}
We relegate the complicated expression for the solution of this differential equation to Appendix~\ref{app_solution}, see Eq.~\eqref{eq:PVM_TC_sol}. In addition, in Appendix~\ref{app:tau} we check the reliability of the adiabatic approximation by comparing its results with a numerical solution of the exact equation of the complete model without the adiabatic elimination. In Fig.~\ref{fig:PVM_tau_deltas}, we plot the time to reach consensus $\tau$ versus the initial condition $\Delta$ for several values of $q$, as obtained in Appendices B and C. For $q>\frac{1}{2}$, the absorbing states are no longer equivalent and therefore $\tau$ is no longer symmetric with respect to its value for $\Delta=0$. As $q$ increases, the time to reach consensus decreases and the maximum of the distribution moves toward the disfavored absorbing state $\Delta=q-1$.

Taking the limit $N \to \infty$ and considering $q_c^-<q<q_c^+$, the analytical expression given by Eq.~\eqref{eq:PVM_TC_sol} simplifies to 
\begin{equation}\label{eq:PVM_TClim}
\lim_{N\to\infty}\tau(\Delta)\sim \mathrm{exp}\left[\frac{\beta}{8}\left(1-\frac{|1-2q|}{\varepsilon}\right)^2\right],
\end{equation}
which means that the initial condition does not matter for large systems and it implies that the time to reach the absorbing state depends exponentially on the system size $N$, since $\beta=\frac{4N\varepsilon^2}{1-\varepsilon^2}$ as shown in the left panel of Fig.~\ref{fig:tau_sizes}. Note that the exponent becomes zero for $q=q_c^{\pm}$. Eq.~\eqref{eq:PVM_TClim} agrees with the result obtained by the WKB theory~\cite{Assaf2010}. In fact, this behavior can be understood as writing the dynamical equation in terms of a gradient in a potential function~\cite{SanMiguel_Toral:2000} $\dot{\Delta}=~-\frac{\partial V}{\partial \Delta}$. For $q_c^-<q<q_c^+$, the self-centered solution is stable while the consensus absorbing fixed points are unstable. For any initial condition different than the consensus states, the system will approach the self-centered solution and stay around it until a large fluctuation, enough to cross the potential barrier, takes it to the absorbing state. The time required for this to happen follows an exponential Arrhenius law in which the inverse of the system size is a measure of the noise intensity~\cite{gardiner}. On the other hand, for $q>q_c^+$ (resp. $q<q_c^-$) the consensus solution $\Delta=q$ (resp. $\Delta=q-1$) becomes stable and the self-centered fixed point disappears, so that the system approaches the favored absorbing solution in a deterministic way. Based on general arguments, a monotonous decay in the potential yields a logarithmic dependence of the time to reach consensus on the system size, as shown in the right panel of Fig.~\ref{fig:tau_sizes}. This is the same dependence found in the biased voter model~\cite{Czaplicka_2022}. The transition between the exponential and logarithmic regimes occurs through a power law behavior for $q=q_c^+$. As evidenced in the insert of the right panel of Fig.~\ref{fig:tau_sizes}, we find numerically that this power law is consistent with an exponent $\frac{1}{2}$.

\subsubsection{Quasi-stationary probability distribution}
In this subsection, we study the behavior of a finite system in which the self-centered solution exists, i.e. for $q_c^-\le q\le q_c^+$, and assumed to survive for a long time, before reaching the absorbing state. The \textit{quasi-stationary distribution}, formally defined as the conditioned probability $P_{\text{qst}}(\Delta)=\lim_{t\to\infty}P(\Delta ,t|\Delta\ne q, \Delta\ne 1-q)$, captures the long term behavior of a process that has not yet reached the absorbing state~\cite{collet2013quasi}. The quasi-stationary distribution can be analyzed using the method proposed in \cite{nasell}, whose details are given in Appendix~\ref{app:quasi-stationary}. From a numerical point of view, the quasi-stationary distribution is obtained averaging over time (after an initial transient) including only those trajectories that have not reached any of the absorbing states. 

\begin{figure}[t]
\centering
\includegraphics[width = \columnwidth]{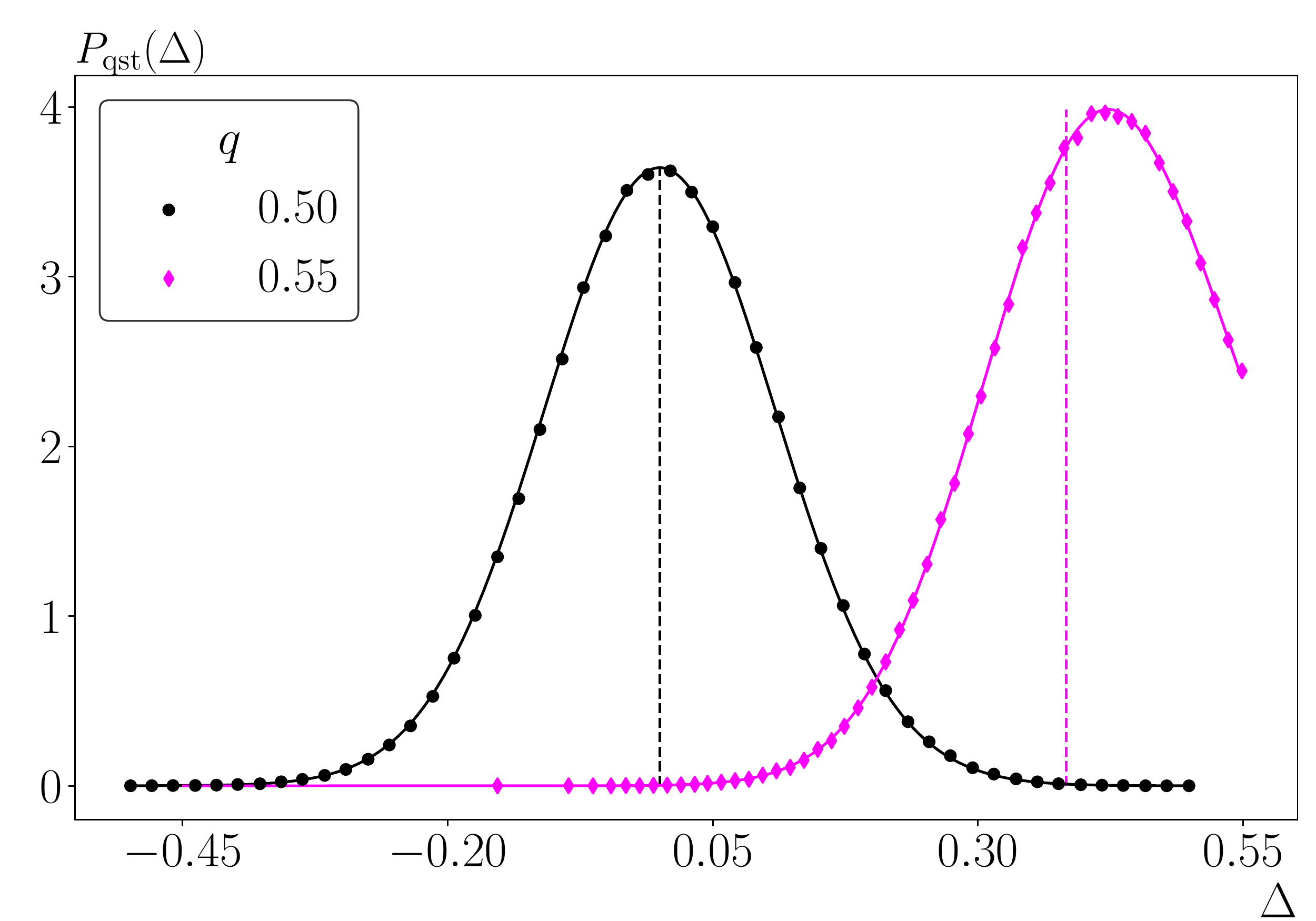}
\caption{Quasi-stationary distribution of $\Delta$. Symbols correspond to computer simulations of the complete model while lines correspond to the solution explained in Appendix~\ref{app:quasi-stationary}. Vertical dashed lines correspond to the value of the self-centered solution $\Delta^*=\frac{1+\varepsilon}{2\varepsilon}\left(2q-1\right)$. Parameter values: $N=1000$, $\varepsilon=0.15$.}
\label{fig:PVM_qsd}
\end{figure}

In Fig.~\ref{fig:PVM_qsd}, we compare the exact solution and the simulations of the complete model. In both symmetric and asymmetric cases the obtained results agree well with the exact solution. In the case $q=0.5$, the quasi-stationary distribution is symmetric around its maximum located at the self-centered fixed point $\Delta=0$. In the asymmetric case $q=0.55$, the distribution is asymmetric around a maximum located at $\Delta\approx 0.393$, a value near, but larger than, the self-centered solution $\Delta^*=\frac{1+\varepsilon}{2\varepsilon}\left(2q-1\right)=0.383\dots$. This is consistent with what we observe in stochastic trajectories, the system tends to the deterministic solution S and remains in its vicinity until a finite-size fluctuation takes it to one of the two absorbing states. 

\section{Noisy Partisan Voter Model}\label{sec:NPVM}
We here introduce the \textit{noisy partisan voter model} including what is known as idiosyncratic changes of state or ``noise''. These changes mimic the mechanism of imperfect imitation by which an agent can, spontaneously and independently from the state of its neighbors, change state at a constant rate $a$. This mechanism has been previously implemented in the standard voter model without preference, resulting in the so-called Kirman's model~\cite{kirman,Artime_2019,CarroSciRep, peraltaNJP}, also known as the \textit{noisy voter model}. The main result of the noisy voter model is that it displays a finite-size noise-induced transition between polarization and consensus for a critical value of the ratio of herding to idiosyncratic rate $\frac{a}{h}$. This critical value depends on the system size $N$ such that it tends to zero as $N$ tends to infinity. The noiseless partisan voter model ($a=0$) studied in the previous chapter depends on the strength of the preference~$\varepsilon$ but it does not present such a transition between polarization and consensus. Our aim in this chapter is to extend the previous study including noise in the form of idiosyncratic changes of opinion and analyze the role of both the preference of the voters $\varepsilon$ and the fraction $q$ of agents that prefer to be in each state in the aforementioned transition.

The stochastic dynamics of the model is modified in the following way: an agent, say $i$, changes its state independently from its neighbors at a constant rate $a$ and selects one of its neighbors at a rate $h$. After selection, agent $i$ follows the dynamic rules explained in Section~\ref{sec:PVM}. Depending on the ratio $\frac{a}{h}$, agents will be more likely to change its state spontaneously or by copying their neighbors. The most relevant change caused by the addition of noise is the disappearance of absorbing states in the system. As in Section~\ref{sec:PVM}, we first make a deterministic analysis of the rate equations and the dynamical system in the mean-field limit before studying the model from the stochastic point of view.

\subsection{Rate equations and dynamical system}
The same set of independent variables, $\Delta$ and $\Sigma$, are used for the analysis. By adding idiosyncratic updates at rate $a$, Eqs.~(\ref{eq:dots1},\ref{eq:dots2}) become
\begin{eqnarray}\label{eq:NPVM_dots1}
\frac{d\Delta}{dt}&=&\varepsilon \Delta - 2\varepsilon \Delta\Sigma+\varepsilon\Sigma\left(2q-1\right)+\frac{2a}{h}\left(2q-1-2\Delta\right), \nonumber \\
&& \\
\frac{d\Sigma}{dt}&=&2 \left(1 + \varepsilon\right)\left(1-q\right) q + \Delta\left(1+2\varepsilon\right)\left(2q-1\right) \nonumber \\ 
&&- \Sigma - 2\varepsilon \Delta^2
+\frac{2a}{h}\left(1-2\Sigma\right),\label{eq:NPVM_dots2}
\end{eqnarray}
where again we have rescaled time conveniently $t\to\frac{h}{2}t$. The determination of the fixed points of this dynamical system leads to a third-degree algebraic equation that can be solved using standard methods. The main feature is that when $a>0$ there is only one single fixed point which is always stable. For a particular value of $q$, the fixed point for $a>0$ corresponds to a shift of the stable fixed point that was present at $a=0$. As the noise intensity $a$ increases, the fixed point moves toward the middle point $(\Delta_0,\Sigma_0)\equiv\left(q-\frac{1}{2},\frac{1}{2}\right)$ corresponding to a situation in which there are the same number of agents in each state. 

For future reference, we write here the nullclines of Eqs.~(\ref{eq:NPVM_dots1}, \ref{eq:NPVM_dots2}) $\dot{\Delta}=~0$ and $\dot{\Sigma}=0$, which are, respectively: 
\begin{eqnarray}
 \label{eq:NPVM_nullclines}
\Sigma &=& \frac{\Delta}{ 1- 2 q + 2 \Delta}+\frac{2a}{h\varepsilon},\\
\Sigma&=&\frac{-2 \Delta^2 \varepsilon + 2 q(1 + \varepsilon) (1 - q) - \Delta (1 + 2 \varepsilon) (1 - 2 q) + \frac{2a}{h}}{1 + \frac{4a}{h}}. \nonumber \\ \label{eq:NPVM_nullcline_sigma}
&& 
\end{eqnarray}

\begin{figure}[t]
\centering
\includegraphics[width=\columnwidth]{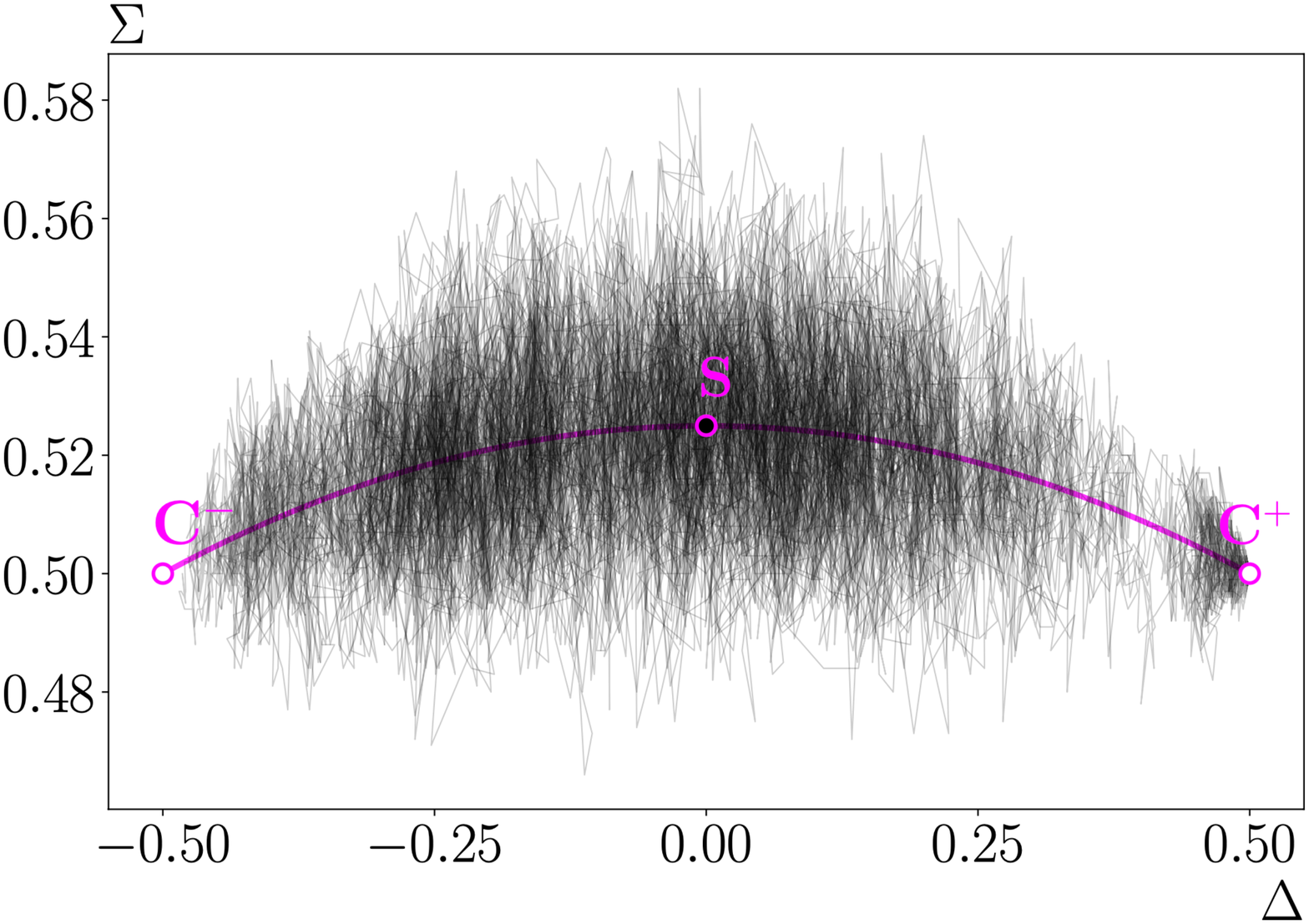}
\caption{Stochastic trajectory in the phase space $(\Delta,\Sigma)$. Darker areas indicate where the system stays for longer time. White dots are the consensus absorbing states. Black dot represents the self-centered solution given by Eqs.~(\ref{eq:NPVM_dots1},\ref{eq:NPVM_dots2}). Magenta line displays the nullcline $\dot{\Sigma}=0$ given by Eq.~\eqref{eq:NPVM_nullcline_sigma}. Parameter values: $N=~1000$, $\varepsilon=0.05$, $q=0.5$, $a=0.00035$, $h=1$.}
\label{fig:NPVM_single}
\end{figure}
\subsection{Stochastic model}

Taking into account the possible interactions between agents in the all-to-all connected topology, the global transition rates in terms of $\Delta$ and $\Sigma$ are given by
\begin{equation} \label{eq:NPVM_globalrates}
\begin{split}
W^{++}&=\frac{Nh}{2}\left[\frac{a}{h}+\left(1-q+\Delta\right)\left(\frac{1+\varepsilon}{2}\right)\right]\left(2q-\Delta-\Sigma\right), \\
W^{+-}&=\frac{Nh}{2}\left[\frac{a}{h}+\left(1-q+\Delta\right)\left(\frac{1-\varepsilon}{2}\right)\right]\left(\Sigma-\Delta\right), \\
W^{-+}&=\frac{Nh}{2}\left[\frac{a}{h}+\left(q-\Delta\right)\left(\frac{1+\varepsilon}{2}\right)\right]\left(2-2q+\Delta-\Sigma\right), \\
W^{--}&=\frac{Nh}{2}\left[\frac{a}{h}+\left(q-\Delta\right)\left(\frac{1-\varepsilon}{2}\right)\right]\left(\Sigma+\Delta\right),
\end{split}
\end{equation}

\begin{figure*}[t] 
\centering
 \includegraphics[width=18 cm]{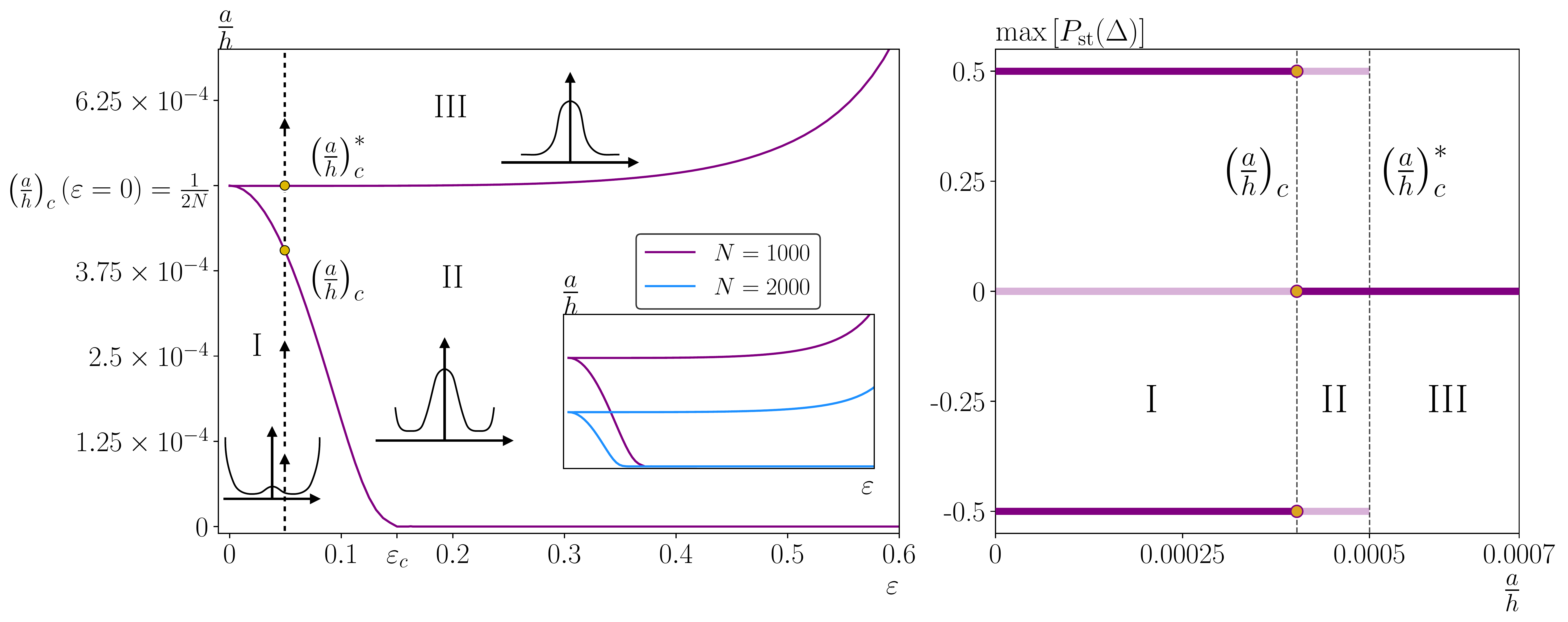}
\caption{Left panel: Parameter diagram $\left(\varepsilon, \frac{a}{h}\right)$ for the different regimes of the stationary probability distribution $P_{\mathrm{st}}(\Delta)$ in the symmetric case $q=\frac{1}{2}$ for a system size $N=1000$. Golden colored dots indicate the transition points between different phases along the path of the dashed line. Inset: Comparison of the parameter diagram $\left(\varepsilon, \frac{a}{h}\right)$ for system sizes: $N=1000$, $2000$ as indicated. Right panel: Location of the maximum/maxima of the stationary probability distribution as a function of the ratio $\frac{a}{h}$ corresponding to the vertical dashed line of the left panel. Light lines indicate the existence of the maximum/maxima while dark lines indicates the absolute maximum/maxima. Golden colored dots indicate where the three maxima are equal. Parameter values: $N=1000$, $\varepsilon=0.05$.}
\label{fig:NPVM_Phase_diagram}
\end{figure*}

from which a master equation for the time evolution of the two-variable probability $P(\Delta,\Sigma;t)$ can be obtained. Again, this master equation is too complicated and we have not been able to extract from it any useful analytical information. However, as we can see in Fig.~\ref{fig:NPVM_single}, despite the noise contribution, the stochastic trajectories still stay around the nullcline $\dot{\Sigma}=0$, Eq.~\eqref{eq:NPVM_nullcline_sigma}. Hence, we use the adiabatic elimination technique and simplify the description using only the slow variable $\Delta$. The resulting drift and diffusion coefficients of Fokker-Planck equation for $P(\Delta,t)$, Eq.~\eqref{eq:FP_1D}, are written down in Appendix~\ref{app:NPVM_FP}. The main difference with the PVM without noise is that, due to the absence of absorbing states, the stationary distribution $P_\text{st}(\Delta)$ can be obtained directly as it does not contain any singularities. $P_\text{st}(\Delta)$ depends on the system size $N$, the fraction $q$ of agents that prefer to be in state $+1$, the preference $\varepsilon$ and the noise intensity $\frac{a}{h}$. In the following, we will analyze the stationary distribution $P_\text{st}(\Delta)$ as the noise increases, fixing $N$ and for different combinations of $q$ and $\varepsilon$, treating separately the symmetric, $q=\frac{1}{2}$, and non-symmetric, $q>1/2$, cases .

\subsubsection{Symmetric case}

The stationary distribution $P_\text{st}(\Delta)$ of the Fokker-Planck equation Eq.~\eqref{eq:FP_1D} obtained from Eqs.~(\ref{eq:Pstform},\ref{eq:potential}) using the drift and diffusion coefficients of Eq.~\eqref{NPVM_FPterms_symm} in the case $q=1/2$ reads
\begin{equation}
\begin{split}\label{eq:NPVM_Pst}
P_{\mathrm{st}}(\Delta)&=\mathcal{Z}^{-1}\mathrm{exp}\left[\frac{-2N\Delta^2\varepsilon^2}{\frac{4a}{h}+1-\varepsilon^2}\right]\\
&\times\left[\left(\frac{a}{h}\right)\left(\frac{4a}{h}+1\right)+\left(\frac{4a}{h}+1-\varepsilon^2\right)\left(\frac{1}{4}-\Delta^2\right)\right]^{\lambda}, \\
\lambda=\frac{a}{h}&\frac{2N}{\left(\frac{4a}{h}+1-\varepsilon^2\right)^2}\left(\frac{4a}{h}+1\right)\left(\frac{4a}{h}+1-2\varepsilon^2\right)-1, 
\end{split}
\end{equation}
where the normalization constant $\mathcal{Z}$ can not be expressed in terms of elementary functions and has to be determined numerically. In the liming case of vanishing preference, $\varepsilon=0$, we recover the known stationary distribution of the noisy voter model which displays a finite-size transition from a bimodal to a unimodal distribution passing through a flat distribution for $\frac{a}{h}=\frac{1}{2N}$~\cite{kirman}. For general $\varepsilon>0$, the system exhibits in the $(\varepsilon,\frac{a}{h})$ parameter space three distinct regimes characterized by a different shape of the stationary distribution. These regimes are separated by two transition lines as shown in Fig.~\ref{fig:NPVM_Phase_diagram}.

\begin{figure*}[t]
    \centering
    \includegraphics[width =\textwidth]{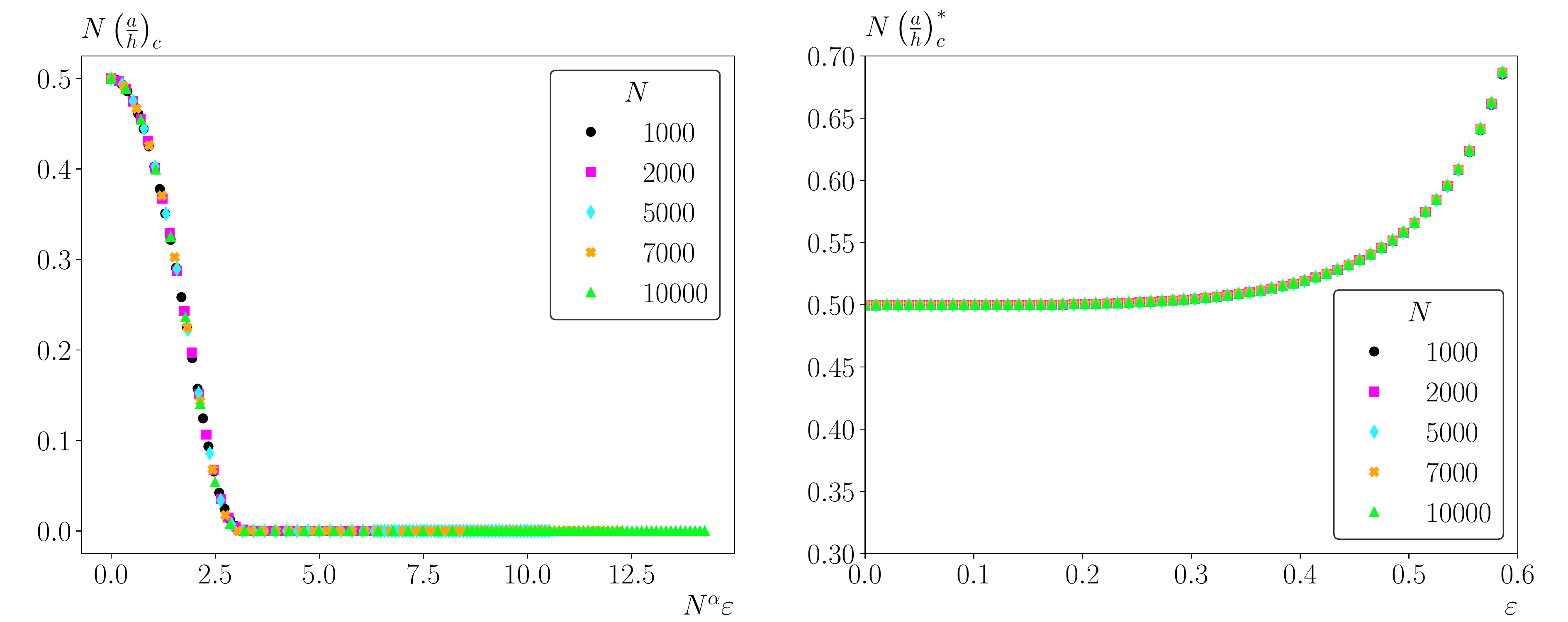}
    \caption{Critical lines $\left(\frac{a}{h}\right)_c$ (left panel) and $\left(\frac{a}{h}\right)^*_c$ (right panel) rescaled by the factor $N$ versus the preference $\varepsilon$ rescaled by $N^{\alpha}$ (left panel) and without rescaling (right panel) for several values of the system size $N$. The best fit has been obtained with an exponent $\alpha=0.439$.}
    \label{fig:NPVM_symmetric_scaling}
\end{figure*}

\begin{itemize}
\item Region I: For $\frac{a}{h}>0$, the distribution is trimodal in this region. The condition for three maxima to exist is that the exponent that appears in Eq.~\eqref{eq:NPVM_Pst} is negative, $\lambda<0$. The maxima, at the boundaries $\Delta=\pm\frac{1}{2}$ and the center $\Delta=0$, are located at the three fixed points of the noiseless dynamics. As $\frac{a}{h}$ or $\varepsilon$ increase, the central peak becomes progressively larger until all three maxima of $P_\mathrm{st}(\Delta)$ are of equal value.

\item \textit{Transition Line} $\left(\frac{a}{h}\right)_c$: It separates \textit{Regions I} and \textit{II} and it describes a discontinuous noise-induced transition. When the system crosses $\left(\frac{a}{h}\right)_c$, there is an abrupt shift of the absolute maximum of the stationary distribution from $\Delta=\pm1/2$ to $\Delta=0$. The transition line intercepts the horizontal axis $\left(\frac{a}{h}\right)_c=0$ at a critical value $\varepsilon_c$, such that for $\varepsilon>\varepsilon_c$ the system avoids region I altogether. This transition line has to be determined numerically and we just plot it in Fig.~\ref{fig:NPVM_Phase_diagram}.

\item Region II: Although the stationary distribution remains trimodal, the central maximum is now larger than the two maxima at the borders. Still it is $\lambda<0$ but, as the ratio $\frac{a}{h}$ increases, the lateral maxima decrease until they disappear at the critical value $\left(\frac{a}{h}\right)_c^*$. 

\item \textit{Transition Line} $\left(\frac{a}{h}\right)_c^*$: This line, defined by the condition $\lambda=0$, corresponds to the boundary between \textit{Regions II} and \textit{III}, indicating the values at which the lateral maxima disappear. When crossing this line from region II, the stationary probability distribution becomes unimodal. 
The transition line $\left(\frac{a}{h}\right)_c^*$ has a complicated analytical expression that can be obtained by standard methods solving the equation $\lambda=0$. Instead we plot the location of the line in Fig.~\ref{fig:NPVM_Phase_diagram}.

\item Region III: The stationary probability distribution is unimodal. As the ratio $\frac{a}{h}$ increases, the maximum at $\Delta=0$ increases its value, such that $P_\text{st}(\Delta)$ becomes a sharper function of its argument.
\end{itemize} 

\begin{figure}[h]
\centering
\includegraphics[width = \columnwidth]{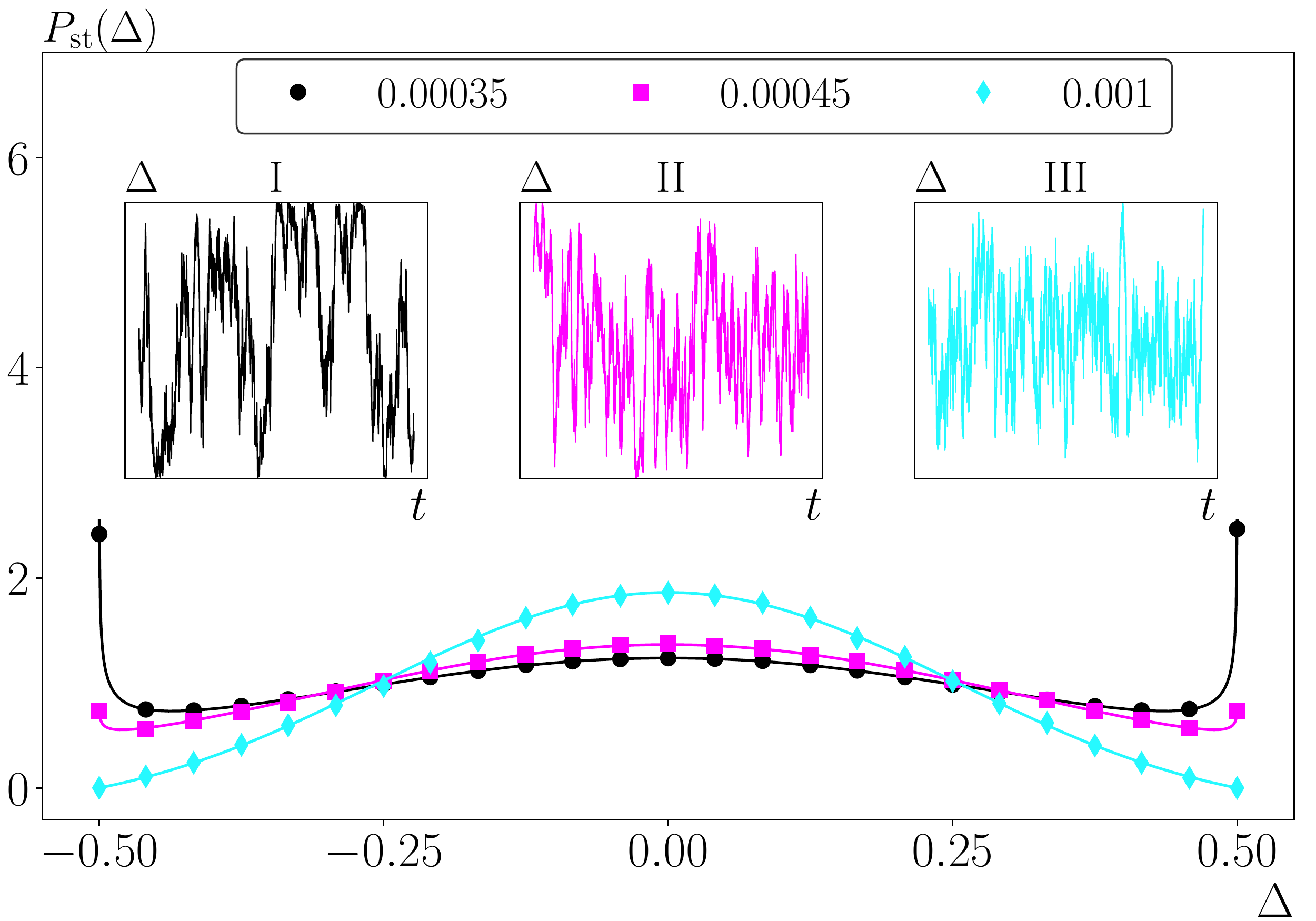} 
\caption{Stationary probability distribution $P_{\mathrm{st}}(\Delta)$ for $\frac{a}{h}=~0.00035, 0.00045, 0.001$, as indicated, in the symmetric case $q=1/2$. Symbols correspond to computer simulations of the complete model while solid lines display the solution given by Eq.~\eqref{eq:NPVM_Pst}. Insets: typical trajectories in each region. The range of values of $\Delta$ in the insets is always $\left[-\frac{1}{2},\frac{1}{2}\right]$. Parameter values: $N=1000$, $\varepsilon=0.05$.}
\label{fig:NPVM_Psts}
\end{figure}

The inset of Fig.~\ref{fig:NPVM_Phase_diagram} demonstrates that both transition lines decrease its value as the system size increases, indicating that they are finite-size transitions. In fact, we have found empirically that the two transition lines can be fitted to the scaling forms
\begin{equation} \label{eq:NPVM_scaling}
\begin{split}
    \left(\frac{a}{h}\right)_c=&N^{-1}\Phi\left(N^{\alpha}\varepsilon\right), \\ 
    \left(\frac{a}{h}\right)^*_c=&N^{-1}\Phi^*\left(\varepsilon\right),
\end{split}
\end{equation}
with an exponent $\alpha=0.439$ determined numerically. To show the validity of this scaling relation we show in  Fig.~\ref{fig:NPVM_symmetric_scaling} a collapse of the data for different values of the system size $N$. In the thermodynamic limit $N\to\infty$, regions I and II disappear and the system only presents a unimodal distribution located at the center of the interval $\Delta=0$. Additionally, the stationary probability distribution in each region together with representative trajectories of the dynamics are displayed in Fig.~\ref{fig:NPVM_Psts}. 

\subsubsection{Asymmetric case}
\begin{figure*}[t] 
\centering
 \includegraphics[width=\textwidth]{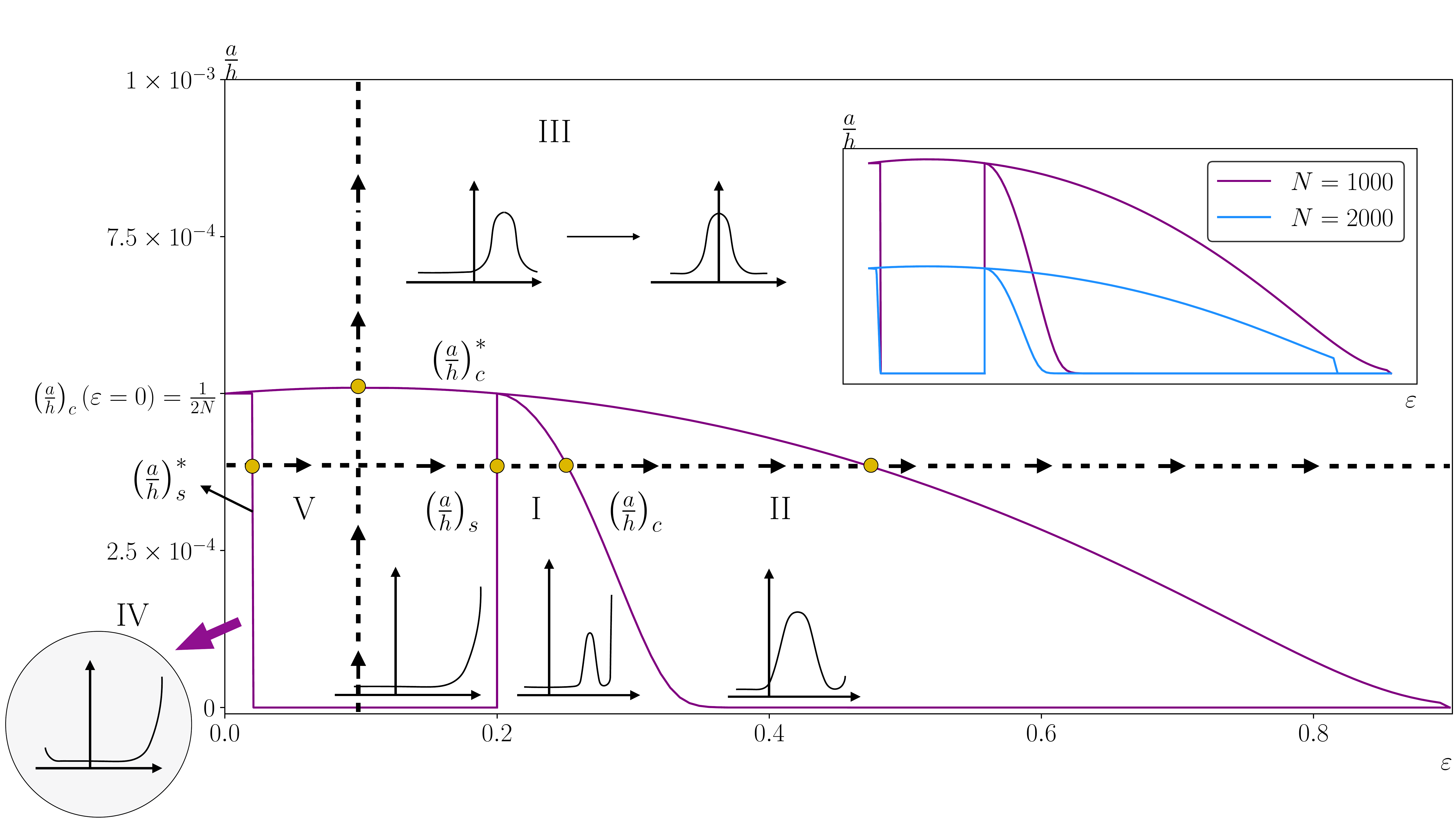}
\caption{Left panel: Parameter diagram $\left(\varepsilon,\frac{a}{h}\right)$ indicating the different regimes of the stationary probability distribution $P_{\mathrm{st}}(\Delta)$. The vertical (resp. horizontal) dashed line display two paths, one increasing $\frac{a}{h}$ and another increasing $\varepsilon$, that have been used in Figs.~\ref{fig:NVPM_asymmetric_pst} and \ref{fig:NPVM_Asymmetric_maxes}. Golden colored dots indicate the transition points along the path of the dashed lines. Parameter values: $N=1000$, $q=0.6$. Inset: Comparison of the transition lines for system sizes: $N=1000$, $2000$ as indicated.}
\label{fig:ANPVM_Phase_diagram}
\end{figure*}

\begin{figure*}
\centering
\includegraphics[width = \textwidth]{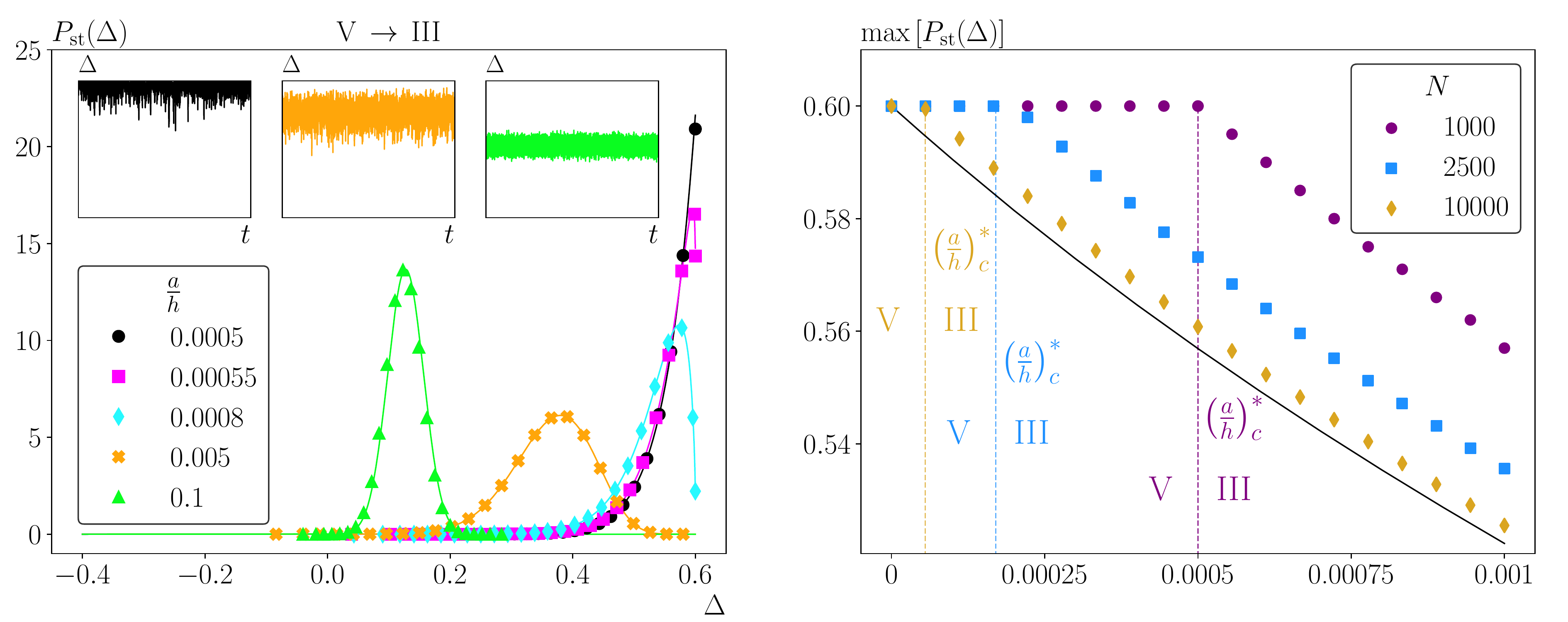}
\caption{Left panel: Stationary probability distributions as a function of $\Delta$ along the vertical dashed line of Fig.~\ref{fig:ANPVM_Phase_diagram}. Symbols correspond to computer simulations of the complete model while lines are the stationary solution of the Fokker-Planck equation given by Eqs.~(\ref{eq:Pstform},\ref{eq:potential}) with the drift and diffusion coefficients of Eq.~\eqref{NPVM_FPterms_symm}. Parameter values: $N=1000$, $\frac{a}{h}=~0.0005, 0.00055, 0.0008, 0.005, 0.1$. 
Insets: Typical trajectories of the dynamics. The range of values of $\Delta$ in the insets is always $\left[-1+q,q\right]$. Right panel: Location of the maximum of the stationary probability distribution versus $\frac{a}{h}$ along the vertical dashed line Fig.~\ref{fig:ANPVM_Phase_diagram} for different system sizes $N=1000, 2500, 10000$ as indicated. Solid black line indicates the position of the stable fixed point solution of Eqs.~(\ref{eq:NPVM_dots1}, \ref{eq:NPVM_dots2}). Common parameter values: $q=0.6$, $\varepsilon=0.1$. }
\label{fig:NVPM_asymmetric_pst}
\end{figure*}

The explicit solution for the stationary distribution $P_\text{st}(\Delta)$ obtained from Eqs.~(\ref{eq:Pstform},\ref{eq:potential}) using the drift and diffusion coefficients of Eq.~\eqref{NPVM_FPterms_symm} for a general value of $q$ is too cumbersome and will not be presented here. In Fig.~\ref{fig:ANPVM_Phase_diagram}, we display the different regions of the parameter diagram $\left(\frac{a}{h},\varepsilon\right)$ for fixed $q$ according to the different shapes that $P_\text{st}(\Delta)$ might take. All our subsequent discussion will consider only the case $q>1/2$, and, for concreteness, in Fig.~\ref{fig:ANPVM_Phase_diagram} we have used $q=0.6$. Let us now describe in detail the different regions and possible transitions amongst them.

\begin{itemize}
\item Region IV: The stationary probability distribution is bimodal. The maximum corresponding to consensus in the preferred state, $\Delta=q$, is much larger than the maximum at the non-preferred state, $\Delta=-1+q$. As $\frac{a}{h}$ or $\varepsilon$ increase, the smaller maximum decreases even further until it disappears at the transition line $\left(\frac{a}{h}\right)_s^{*}$ into region V. Although not visible at the scale of the figure, except when moving along the line $\varepsilon=0$, there is no intersection between regions IV and III, and a path leading from IV to III must pass necessarily through region V.

\item Region V: The stationary distribution is unimodal. The maximum is located at the preferred state $\Delta=q$. For fixed $\varepsilon$, an increase of $\frac{a}{h}$ beyond the value $\left(\frac{a}{h}\right)_c^{*}$ leads to the transition to region III. Alternatively, for fixed $\frac{a}{h}$, an increase of $\varepsilon$ beyond the transition line labeled $\left(\frac{a}{h}\right)_s$, and that corresponds to the vertical line determined by $\varepsilon=2q-1$, leads to region I.

\item Region I: The stationary probability distribution is bimodal. The absolute maximum is still located at the boundary $\Delta=q$. When crossing from region V the transition line labeled $\left(\frac{a}{h}\right)_s$, a second maximum begins to appear starting at the same value $\Delta=q$. As $\varepsilon$ or $\frac{a}{h}$ increases, this second maximum moves to the center of the $\Delta$ interval and grows until both maxima become equal in height when the system crosses the transition line $\left(\frac{a}{h}\right)_c$ towards region~II. Except when moving along the line $\varepsilon=2q-1$, there is no path leading directly from region~I to region~III. 

\item Region II: The stationary probability distribution is still bimodal. The larger maximum is located close to the self-centered value $\Delta=\Delta^*(a)$, while the smaller maximum is in the preferred state $\Delta=q$. As $\frac{a}{h}$ or $\varepsilon$ increase the second maximum disappears when crossing the line $\left(\frac{a}{h}\right)_c^*$ towards region III.

\item Region III: The probability distribution is unimodal. This region can be reached directly from regions V and II when crossing the respective transition lines. As $\frac{a}{h}$ or $\varepsilon$ increases within region III, the maximum grows and moves toward the center, $\Delta=\Delta_0$, of the interval of the allowed values for $\Delta\in(-1+q,q)$.
\end{itemize}

Regions I, II and III have a correspondence with the regions of the same name in the symmetric case, but the shape of the probability distribution is modified under the presence of the symmetry breaking condition $q\ne~1/2$. Regardless of the value of $q$, I and II are regions in which the probability distribution is bimodal. The relative maximum that is not at the extreme of the interval reflects the existence of the self-centered solution for $a=0$. Region III displays a single maximum, noise-dominated probability distribution. 
In regions IV and V, there is no self-centered solution and the dominance of the maximum at the consensus solution $\Delta=q$ arises from the symmetry breaking condition $q>\frac{1}{2}$. As shown in the inset of Fig.~\ref{fig:ANPVM_Phase_diagram}, regions IV, V, I and II become smaller as the system size increases. In the thermodynamic limit, only region III persists and the maximum of the stationary probability distribution is located at the corresponding position of the stable fixed point of the rate equations. 

In the left panel of Fig.~\ref{fig:NVPM_asymmetric_pst}, we fix $\varepsilon<2q-1$ and plot the stationary probability distribution, together with representative trajectories, for several values of $\frac{a}{h}$, which corresponds to points along the vertical dashed line of Fig.~\ref{fig:ANPVM_Phase_diagram}. As shown in the right panel of this figure, when crossing the line $\left(\frac{a}{h}\right)_c^{*}$, the system undergoes a continuous noise-induced transition, as indicated by a continuous decrease of the location of the absolute maximum of the probability distribution. The value of transition point $\left(\frac{a}{h}\right)_c^{*}$ decreases as the system size $N$ increases, indicating that this transition will disappear in the thermodynamic limit. Note that the location of the maximum of the probability distribution of region III does not correspond to the stable deterministic solution, but they approach each other as the system size increases. 

\begin{figure*}
\includegraphics[width = \textwidth]{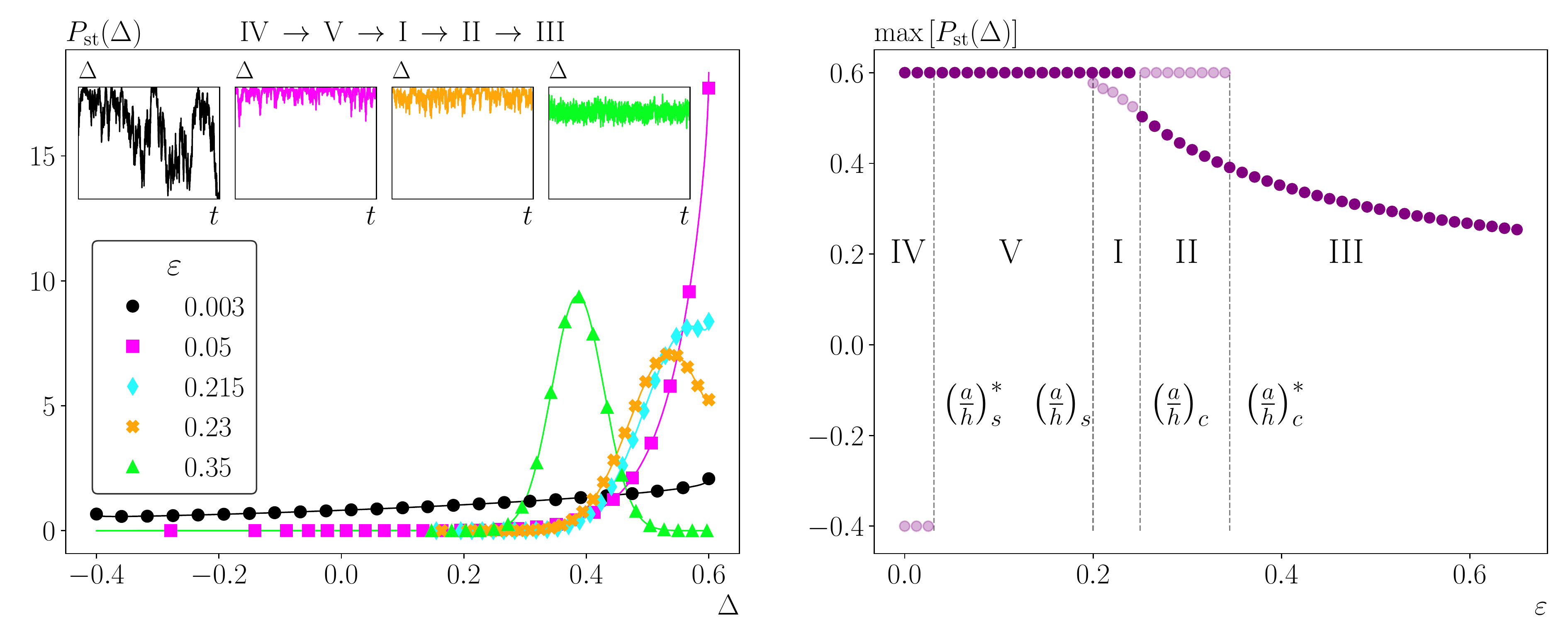}
\caption{Left panel: Stationary probability distributions as a function of $\Delta$ along the horizontal dashed line of Fig.~\ref{fig:ANPVM_Phase_diagram}. Symbols correspond to computer simulations of the complete model while lines are the stationary solution of the Fokker-Planck equation given by Eqs.~(\ref{eq:Pstform},\ref{eq:potential}) with the drift and diffusion coefficients of Eq.~\eqref{NPVM_FPterms_symm}. Parameter values: $\varepsilon=~0.003, 0.05, 0.215, 0.23, 0.35$. Insets: Typical trajectories of the dynamics. The range of values of $\Delta$ in the insets is always $\left[-1+q,q\right]$. Right panel: Location of the maxima of the stationary probability distribution versus $\varepsilon$ along the vertical dashed line Fig.~\ref{fig:ANPVM_Phase_diagram}. Light dots correspond to the existence of a relative maximum while dark dots indicates the absolute maximum. Common parameter values: $N=1000$, $q=0.6$, $\frac{a}{h}=0.0004$.}

 \label{fig:NPVM_Asymmetric_maxes}
\end{figure*}

In the left panel of Fig.~\ref{fig:NPVM_Asymmetric_maxes}, we fix $\frac{a}{h}<\left(\frac{a}{h}\right)_c(\varepsilon=0)$ and plot the stationary probability distribution, together with representative trajectories, for several values of $\varepsilon$, which corresponds to points along the horizontal, dashed line of Fig.~\ref{fig:ANPVM_Phase_diagram}, passing through regions IV$\to$V$\to$I$\to$II. In the right panel of that figure, we plot the maxima of the probability distribution as a function of $\varepsilon$ when moving along this line. Most noticeably, the transition from regions I and II, when crossing the line $\left(\frac{a}{h}\right)_c$, is discontinuous as there is an abrupt shift of the absolute maximum of the probability distribution, while in the transitions from IV to V and from V to I the location of the absolute maximum does not vary.

\section{Summary and conclusions}\label{summary}
In this paper we have studied the impact of including preference for one of the two states to the agents in both the voter model and the noisy voter model. Regarding the Partisan Voter Model, we have revisited the model from the deterministic point of view and we present a thorough stochastic analysis of the model. We have used an adiabatic elimination technique by taking advantage of the different time scale of the variables. The results of both the exit probability and fixation time derived using the adiabatic approximation have been compared with those obtained with the complete model substantiating the validity of this approximation for $\varepsilon\ll1$. For a finite system, the stationary probability distribution $P_{\mathrm{st}}(\Delta)$ has two delta functions in the absorbing states $\Delta=q,\,1-q$ weighted with the probability to reach one of them. We have found an analytical expression for the exit probability. For large $N$, the system ends up in any absorbing state with equiprobability if $q\neq\frac{1}{2}$ or in the favored state otherwise. For a finite system, one of the two absorbing states is reached in a finite time $\tau$. In addition, for large enough systems we demonstrate that the fixation time exponentially depends on the system size. Finally, the system exhibits constant probability distribution before reaching an absorbing state, i.e. a quasi-stationary distribution, whose maximum is located at the self-centered fixed point, if it exists, i.e. $q_c^-<q<q_c^+$. 

Regarding the Noisy Partisan Voter Model, the description of the system has been simplified again with the adiabatic elimination technique. Using this approximation we have found the transition lines for the noise induced transitions that occur in this model and that differ in non-trivial ways from those that appear in the ordinary Noisy Voter model. 
On the one hand, in the symmetric case, we have found three different regions separated by two transition lines in the parameter space $\left(\varepsilon,\frac{a}{h}\right)$ based on the shape of the stationary probability distribution $P_{\mathrm{st}}(\Delta)$. The steady distribution presents a maximum located at the self-centered fixed point. There is a discontinuous transition from a trimodal to a unimodal distribution. The transition is finite-sized and it disappears in the thermodynamic limit. On the other hand, the asymmetric case enriches the parameter diagram by adding two more regions and modifying the three previous ones. If the self-centered solution does not exist in the limit $a=0$, the system exhibits a bimodal distribution wherein the relative maximum diminishes while the absolute maximum undergoes a continuous transition from the favored absorbing state towards the center of the $\Delta$ space. If the self-centered solution exists in the limit $a=~0$, the steady distribution is also bimodal. In this case the lower maximum is located close to the self-centered solution. The system undergoes a discontinuous noise-induced transition by increasing either $\frac{a}{h}$ or $\varepsilon$. After the disappearance of the lateral maximum, the absolute maximum moves continuously towards the center of the interval. In the thermodynamic limit, all transitions disappear and the system presents a unimodal distribution centered at the stable fixed point of the rate equations, Eqs.~(\ref{eq:NPVM_dots1}, \ref{eq:NPVM_dots2}).

Future work includes the study of the model on regular and random networks, as well as the inclusion of non-linear interactions which are known to transform a finite-size noise-induced transition into a bona-fide phase transition present in the thermodynamic limit \cite{PeraltaCHAOS,Artime_2019}.\\

\acknowledgments{
Partial financial support has been received from the Agencia Estatal de Investigaci\'on (AEI, MCI, Spain) MCIN/AEI/10.13039/501100011033 and Fondo Europeo de Desarrollo Regional (FEDER, UE) under Project APASOS (PID2021-122256NB-C21) and the María de Maeztu Program for units of Excellence in R\&D, grant CEX2021-001164-M.}

\bibliography{references}

\clearpage

\appendix
\setcounter{figure}{0}
\renewcommand{\thefigure}{A\arabic{figure}}
\widetext

\section{Calculation of the exit probability}\label{app:PVM_exitp_complete}
Let $P_q\left(\Delta,\Sigma\right)$ be the probability of reaching the absorbing state $(\Delta=q,\Sigma=q)$ starting from an initial condition $\left(\Delta_0=\Delta,\Sigma_0=\Sigma\right)$. This probability is related to the possible transitions as \cite{redner_2001}
\begin{equation} \label{app:eq:PVM_exitp_complete}
P_q\left(\Delta,\Sigma\right)={\cal L}\left[P_q\left(\Delta,\Sigma\right)\right],
\end{equation}
where we have defined the linear operator
\begin{equation} \label{app:eq:PVM_exitp_complete2}
\begin{split}
{\cal L}\left[P_q\left(\Delta,\Sigma\right)\right]=
&\quad w^{++}(\Delta,\Sigma)P_q\left(\Delta+1/N, \Sigma+1/N\right)
+w^{+-}(\Delta, \Sigma)P_q\left(\Delta+1/N, \Sigma-1/N\right)\\
&+w^{-+}(\Delta, \Sigma)P_q\left(\Delta-1/N, \Sigma+1/N\right)
+w^{--}(\Delta, \Sigma)P_q\left(\Delta-1/N, \Sigma-1/N\right),
\end{split}
\end{equation}
being 
\begin{equation}
\begin{split}\label{eq:PVM_globalrates_rescaled}
 w^{\pm,\pm}(\Delta, \Sigma)&=\frac{W^{\pm,\pm}(\Delta, \Sigma)}{\Omega(\Delta,\Sigma)},\\
 \Omega(\Delta,\Sigma)&=W^{++}(\Delta,\Sigma)+ W^{+-}(\Delta,\Sigma)+W^{-+}(\Delta,\Sigma)+ W^{--}(\Delta,\Sigma)
 \end{split}
\end{equation}
and the global rates $W^{\pm}(\Delta,\Sigma)$ are given by Eqs.~\eqref{eq:PVM_globalrates}. The recurrence relation of Eq.~\eqref{app:eq:PVM_exitp_complete} has to be implemented with the boundary conditions
\begin{equation}\label{eq:boundary}
\begin{split}
&P_q(\Delta=-1+q, \Sigma=1-q)=0,\\
&P_q(\Delta=q,\Sigma=q)=1. 
 \end{split}
\end{equation}
Unfortunately, we have not been able to find the solution of Eq.~\eqref{app:eq:PVM_exitp_complete} in a closed form. Instead, we have used a simple iteration procedure in which an initial guess for the probabilities $P_q^{(0)}(\Delta,\Sigma)$ is iterated
\begin{equation} \label{app:eq:exitp_recurrence}
P_q^{(k+1)}(\Delta,\Sigma)={\cal L}\left[P_q^{(k)}\left(\Delta,\Sigma\right)\right],\quad k=0,1,\dots
\end{equation}
until convergence to the solution $P_q(\Delta,\Sigma)$. Note that the boundary conditions Eq.~\eqref{eq:boundary} need to be imposed at every iteration. We consider that convergence has been reached when $\displaystyle\left|\sum_{\Delta,\Sigma}\left[P^{(k+1)}(\Delta,\Sigma)-P^{(k)}(\Delta,\Sigma)\right]\right|<10^{-8}$. This simple procedure works well for small values of $N$, but it becomes prohibitive for large $N$, as it turns out that the CPU time needed to reach convergence scales roughly as $N^4$. Otherwise, the obtained values are precise within the numerical accuracy of the iteration procedure. 

\begin{figure*}
\centering
\includegraphics[width = \textwidth]{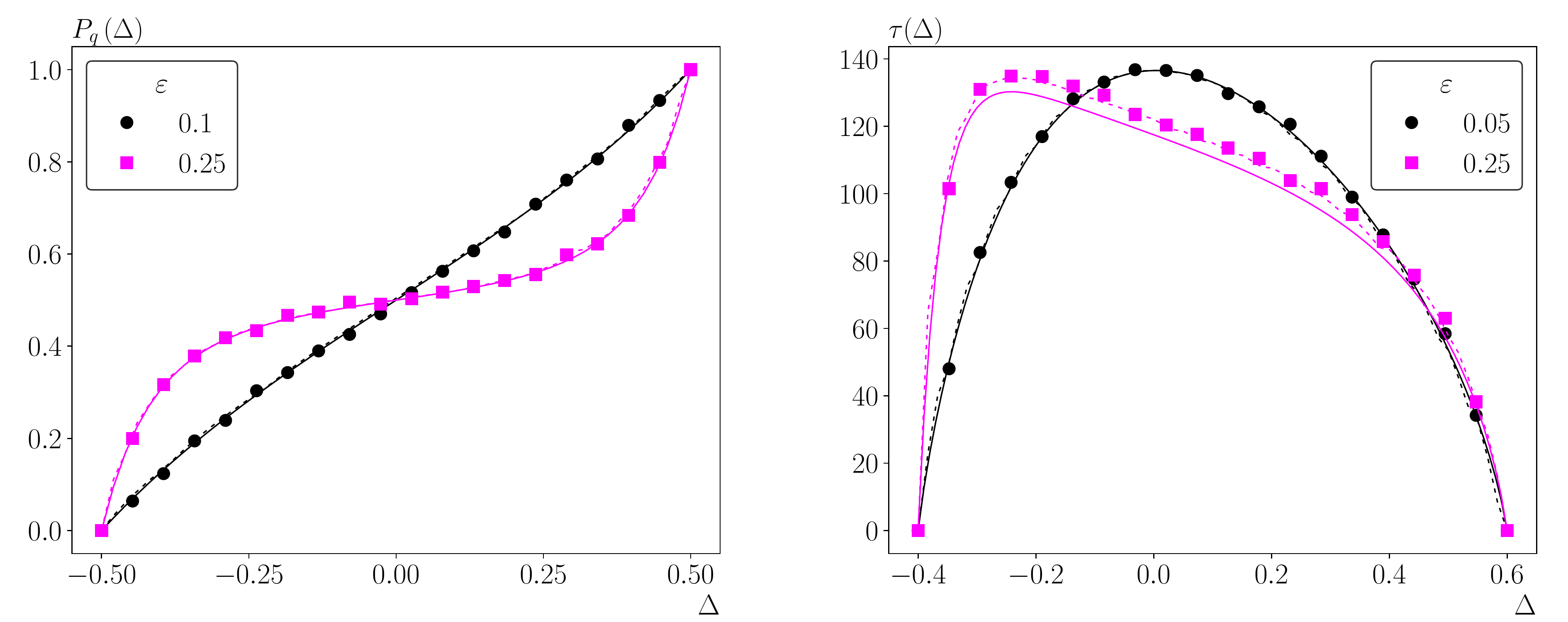}
\caption{Left panel: Exit probability $P_q$ versus the initial condition $\Delta$. Parameter value: $q=0.50$. Right panel: Time to reach an absorbing state $\tau$ as a function of the initial condition $\Delta$. Parameter value: $q=0.60$. In both panels, the system size is $N=100$. Solid lines represent the solution given by the adiabatic approximation, Eqs.~(\ref{eq:PVM_PexitQ}, \ref{eq:PVM_TC_sol}), respectively, while symbols correspond to computer simulations of the complete model and dashed lines display the solution of the recursion relation Eqs.~(\ref{app:eq:exitp_recurrence}, \ref{app:eq:PVM_tau_complete_iter}), respectively. In the latter two cases we have taken the value of $\Sigma$ at the nullcline, Eq.~\eqref{eq:PVM_nullcline_sigma}.}
\label{app:fig:tau}
\end{figure*}

In left panel of Fig.~\ref{app:fig:tau} we compare the adiabatic approximation solution Eq.~\eqref{eq:PVM_PexitQ} and the result obtained with this numerical method for $P_q(\Delta,\Sigma)$ as a function of $\Delta$ taking for $\Sigma$ the value at the nullcline Eq.~\eqref{eq:PVM_nullcline_sigma}. However, it should be noted that the discrete grid points $(\Delta,\Sigma)$ do not, in general, lie exactly on the nullcline. Therefore, each value is computed based on a weighted average of the four closest grid locations. The weight assigned to each point is the inverse of its Euclidean distance to the coordinates $(\Delta,\Sigma)$. As shown in the figure, the agreement between theory and simulations is very good for the displayed values of $\varepsilon$ and, in fact, we have observed that the agreement is satisfactory for $\varepsilon\lesssim 0.5$. 

\section{Solution of Eq.~\eqref{eq:PVM_TC_aprr}}
\label{app_solution}
Due to the linearity of Eq.~\eqref{eq:PVM_TC_aprr} and the symmetry under the change $\Delta \to -\Delta$ and $q \to 1-q$ , the solution of the differential equation can be expressed as
\begin{equation} \label{eq:PVM_tau_sol}
\tau\left(\Delta\right)=\frac{2 N}{{1-\varepsilon^2}}\left[T\left(\Delta;q\right)+T\left(-\Delta;1-q\right)\right],
\end{equation}
where $T\left(\Delta;q\right)$ fulfills
\begin{equation}\label{eq:PVM_TC_part}
\frac{d^2T}{d\Delta^2}=\left(\beta \Delta+\gamma\right)\frac{dT}{d\Delta}-\frac{1}{\left(q-\Delta\right)},
\end{equation}
with boundary conditions $T\left(\Delta=q\right)=T\left(\Delta=q-1\right)=0$.

The general solution of this linear differential equation can be found by standard methods in terms of two integration constants that are determined using the boundary conditions. Unfortunately, in this case, the integrals required by the solution to the equation cannot be expressed using elementary functions. Instead, we present the solution as 
\begin{equation}\label{eq:PVM_TC_sol}
\begin{split}
T\left(\Delta\right)&=\sqrt{\frac{\pi}{2\beta}} \left( C\left(\beta,\gamma\right)\left[\mathrm{erfi}\left(\sqrt{\frac{\beta}{2}}\left(\Delta+\frac{\gamma}{\beta}\right)\right) -\mathrm{erfi}\left(\sqrt{\frac{\beta}{2}}\left(q-1+\frac{\gamma}{\beta}\right)\right)\right] \right.\\[8pt]
& \left.-\bigintss_{q-1}^{\Delta}dx\,\frac{\exp{\left[-\frac{\beta}{2}\left(x+\frac{\gamma}{\beta}\right)^2\right]}}{q-x}\left[\mathrm{erfi}\left(\sqrt{\frac{\beta}{2}}\left(\Delta+\frac{\gamma}{\beta}\right)\right)-\mathrm{erfi}\left(\sqrt{\frac{\beta}{2}}\left(x+\frac{\gamma}{\beta}\right)\right)\right] \right), \\[8pt]
C(\beta,\gamma)&=\frac{\bigintss_{q-1}^{q}dx\,\frac{\exp{\left[-\frac{\beta}{2}\left(x+\frac{\gamma}{\beta}\right)^2\right]}}{q-x}\left[\mathrm{erfi}\left(\sqrt{\frac{\beta}{2}}\left(q+\frac{\gamma}{\beta}\right)\right)-\mathrm{erfi}\left(\sqrt{\frac{\beta}{2}}\left(x+\frac{\gamma}{\beta}\right)\right)\right]}{\mathrm{erfi}\left(\sqrt{\frac{\beta}{2}}\left(q+\frac{\gamma}{\beta}\right)\right)-\mathrm{erfi}\left(\sqrt{\frac{\beta}{2}}\left(q-1+\frac{\gamma}{\beta}\right)\right)},
\end{split}
\end{equation}
which requires the numerical determination of two integrals. 

\section{Calculation of the time to reach consensus}\label{app:tau}
Let $\tau\left(\Delta,\Sigma\right)$ be the average time to reach any absorbing state starting from an initial condition $\left(\Delta_0=\Delta,\Sigma_0=\Sigma\right)$. This time is related to the 
possible transitions as \cite{redner_2001}
\begin{equation} \label{app:eq:PVM_tau_complete}
\begin{split}
\tau\left(\Delta,\Sigma\right)=\frac{1}{\Omega(\Delta,\Sigma)}+{\cal L}\left[\tau\left(\Delta,\Sigma\right)\right],
\end{split}
\end{equation}
where the operator $\cal L$ is defined in Eq.~\eqref{app:eq:PVM_exitp_complete2}, and $\Omega(\Delta,\Sigma)$ in Eq.~\eqref{eq:PVM_globalrates_rescaled}, together with the boundary conditions
\begin{equation}\label{eq:boundary_tau}
\begin{split}
&\tau(\Delta=-1+q, \Sigma=1-q)=0,\\
&\tau(\Delta=q,\Sigma=q)=0.
\end{split}
\end{equation}

Again we solve numerically Eq.~\eqref{app:eq:PVM_tau_complete} using an iteration scheme
\begin{equation} \label{app:eq:PVM_tau_complete_iter}
\begin{split}
\tau^{(k+1)}\left(\Delta,\Sigma\right)=\frac{1}{\Omega(\Delta,\Sigma)}+{\cal L}\left[\tau^{(k)}\left(\Delta,\Sigma\right)\right],
\end{split}
\end{equation}
starting from an initial guess $\tau^{(0)}(\Delta,\Sigma)$ and iterating until convergence, imposing the boundary conditions Eq.~\eqref{eq:boundary_tau} at each iteration. Convergence is determined by the condition $\displaystyle\left|\sum_{\Delta,\Sigma}\left[\tau^{(k+1)}(\Delta,\Sigma)-\tau^{(k)}(\Delta,\Sigma)\right]\right|<10^{-8}$. As in the case of the exit probability, this numerical procedure is very precise but it is only feasible for relatively small values of $N$ because of the large CPU time needed for convergence.

In the right panel of Fig.~\ref{app:fig:tau} we compare the solution obtained with this method and the solution given by Eq.~\eqref{app:eq:PVM_tau_complete_iter}, which includes the adiabatic approximation. Note that, unlike the exit probability, the adiabatic approximation performs better for small values of $\varepsilon$, while there are systematic discrepancies for larger values.

\section{Calculation of the quasi-stationary probability distribution}
\label{app:quasi-stationary}
We here calculate numerically the quasi-stationary probability distribution of $\Delta$, along the lines of the method exposed in \cite{nasell}.

For practical purposes we introduce the index $j=N(\Delta+1-q)$ which takes integer values in the interval $0\le j\le N$, and we write $P_j(t)\equiv P\left(\Delta=\frac{j}{N}+q-1,t\right)$.
The master equation of the process in the adiabatic approximation regime is given by~\cite{Toral2014StochasticNM} 
\begin{equation}\label{app:eq:mastereqADB}
\frac{\partial P_j(t)}{\partial t}= \left( E^{-1} - 1\right)\left[ W^{+}_j P_j(t)\right] +\left( E - 1\right) \left[ W^{-}_j P_j(t)\right],
\end{equation}
where $E$ is the step operator acting on any function $f(j)$ as $E^{\ell}[f(j)]=f(j+\ell)$. The global rates can be derived from those of Eqs.~\eqref{eq:reducedglobalrates} as
\begin{equation} \label{app:eq:reducedglobalrates}
\begin{split}
W^+_j&=\frac{j}{2} \left(1-\frac{j}{N}\right)
 \left(1-2 \varepsilon^2\frac{ j}{N}-\varepsilon \left(1-2q\right)\right), \\
W^-_j& =\frac{j}{2} \left(1-\frac{j}{N}\right)
 \left(1-2\varepsilon^2
 \left(1-\frac{j}{N}\right)+\varepsilon\left(
 1-2q\right)\right).
\end{split}
\end{equation}
Note that both vanish at the absorbing states $j=0$ and $j=N$. 

We now define the probability $Q_j(t)$ of the system to be in the state $j$ at time $t$, \textit{conditioned to non having reached the absorbing state}, i.e. $Q_j(t)=\mathrm{Prob}(j|j\neq 0, N;t)$, as
\begin{equation} \label{app:eq:Q}
Q_j(t)=\frac{P_j(t)}{1-P_0(t)-P_N(t)}, \hspace{10pt} j=1, \dots,N-1.
\end{equation}
\vspace{0.1cm}

Differentiating this relation and using 
\begin{equation}
\begin{split}
\frac{\partial P_0(t)}{\partial t}&=W^-_1P_1(t), \\
\frac{\partial P_N(t)}{\partial t}&=W^+_{N-1}P_{N-1}(t),
\end{split}
\end{equation}
obtained from Eq.~\eqref{app:eq:mastereqADB}, we can write down the master equation of the process conditioned to non having reached the absorbing state as
\begin{equation}\label{eq_qst}
\frac{\partial Q_j(t)}{\partial t}=\left( E^{-1} - 1\right)\left[ W^{+}_j Q_j(t)\right] +\left( E - 1\right) \left[ W^{-}_j Q_j(t)\right]+Q_j(t)(W_1^-Q_1(t) +W_{N-1}^+ Q_{N-1}(t))\\
\end{equation}
with the convention $W_0^+Q_0=W_N^-Q_N=0$. The quasi-stationary distribution $Q_j$ is defined as the stationary solution of this master equation obtained by setting the time derivative equal to zero. Alternatively, we can also consider the long time solution $Q_j=\lim_{t\to\infty}Q_j(t)$. This has been obtained by integrating numerically the evolution equation \eqref{eq_qst} using the Euler method:
\begin{equation}
Q_j(t+\Delta t)=Q_j(t)+\Delta t \frac{\partial Q_j(t)}{\partial t},\hspace{10pt} j=1, \dots,N-1.
\end{equation}
with an initial condition $Q_j(t=0)=\frac{1}{N-1},\, \forall j$. The procedure is iterated until there is no significant change in the value of $Q_j(t)$. Note that the normalization condition $\sum_{j=1}^{N-1}Q_j(t)=1$ is exactly satisfied by the Euler method at all times and no special requirement is needed for the time step $\Delta t$, other that it does not lead to a numerical instability. We have found that $\Delta t=0.001$ is a convenient value.
In Fig.~\ref{fig:PVM_qsd} we compare this numerical solution with the results of computer simulations showing a good agreement between both.

\section{Fokker Planck equation for the NPVM}
\label{app:NPVM_FP}

The effective rates for this variable can be derived from those of Eq.~\eqref{eq:NPVM_globalrates} as

\begin{equation}
\begin{split}
W^+(\Delta)&=\frac{Nh}{2}\left[\frac{2a}{h}\left(q-\Delta\right)+\frac{1-q+\Delta}{1+\frac{4a}{h}}\left(\left(\Delta-q\right)\left(\varepsilon\left(1-2q\right)-1+2\varepsilon^2(1-q+\Delta)\right)-\frac{2a}{h}\left(2\Delta+\varepsilon-2q(1+\varepsilon)\right)\right)\right],\\
W^-(\Delta)&=\frac{Nh}{2}\left[\frac{2a}{h}\left(1-q+\Delta\right)-\frac{\Delta-q}{1+\frac{4a}{h}}\left(\left(1+\Delta-q\right)\left(\left(1+\varepsilon\right)\left(1-2\varepsilon q\right)+2\Delta\varepsilon^2\right)+\frac{2a}{h}\left(1+2\Delta+\left(1+\varepsilon\right)\left(1-2 q\right)\right)\right)\right].
\end{split}
\end{equation}

These global rates lead us to the Fokker-Planck equation \eqref{eq:FP_1D} of the noisy version of the partisan voter model where the drift and diffusion terms are given by

\begin{equation}\label{NPVM_FPterms_symm}
\begin{split}
F(\Delta)&=\frac{h}{1+\frac{4a}{h}}\left[\varepsilon \left(\Delta-q\right)\left(1+\Delta-q\right)\left(2\Delta\varepsilon+\left(1-2q\right)(1+\varepsilon)\right)-\frac{a}{h}\left(2\Delta+\left(1-2q\right)(1+\varepsilon)\right)-4\left(\frac{a}{h}\right)^2\left(2\Delta+1-2q\right)\right], \\
D(\Delta)&=\frac{h}{2N}\frac{1}{1+\frac{4a}{h}}\left[-\Delta^2\left(1-\varepsilon^2+\frac{4a}{h}\right)+\left(1-\varepsilon+\frac{4a}{h}\right)\left(\frac{a}{h}+q(1-q)(1+\varepsilon)\right)-\Delta\left(1-2q\right)\left(1-\varepsilon^2+(2+\varepsilon)\frac{2a}{h}\right)\right],
\end{split}
\end{equation}
respectively. The stationary probability distribution $P_{\mathrm{st}}(\Delta)$ can be obtained with Eqs.~(\ref{eq:Pstform}, \ref{eq:potential}), although the integrals can be performed analytically using symbolic manipulation programs such as Mathematica~\cite{Mathematica}, the resulting expression is too long and will not be reproduced here.

\end{document}